# Graphene Nanobubble: A New Optical Nonlinear Material

Qiaoliang Bao[1,2]*, Jianqiang Chen[3,4]*, Yuanjiang Xiang[1], Kai Zhang[1], Shaojuan Li[2], Xiaofang Jiang[1], Qing-Hua Xu[1], Kian Ping Loh[1]†, T. Venkatesan[3,4,5]†

[1]Department of Chemistry, and Graphene Research Centre, National University of Singapore, 3 Science Drive 3, 117543, Singapore.

[2]Institute of Functional Nano & Soft Materials (FUNSOM), Soochow University, Suzhou, Jiangsu 215123, China.

[3]Department of Electrical and Computer Engineering, National University of Singapore, 117576, Singapore.

[4]NUSNNI-NanoCore, National University of Singapore, 117576, Singapore.

[5]Department of Physics, National University of Singapore, 117576, Singapore.

*These authors contributed equally to this work.

†To whom correspondence should be addressed. E-mail: chmlohkp@nus.edu.sg (K. P. L.); venky@nus.edu.sg (T. V.)

# Abstract

Graphene is a rising star in nonlinear optics due to its saturable absorption and giant Kerr nonlinearity, these properties are useful in digital optics based on optical nonlinear devices. However, practical applications require large optical nonlinearities and these are inherently limited by the interaction length of atomically thin graphene. Here, we demonstrate optical bistability in a Fabry–Perot cavity containing monolayer and bilayer graphene which have been restructured to form nanobubbles. We find that graphene nanobubble can act as a new type of optical nonlinear media due to its vertical side wall as well as added curvature, which enable strong non-linear dispersive effects leading to a large optically induced phase change. Unlike thermally induced bistability, the all-optical switching between two transmission states happens within a time scale of tens of nanoseconds. Nanobubble-based optical devices with intrinsic optical nonlinearity help to overcome the optical path length limitation of atomically thin two dimensional films and allow us to explore the promise of using such elements as the building block of digital all-optical circuitry.

Graphene photonics offers an unprecedented opportunity to study light-matter interaction involving relativistic particles in an ultrathin two-dimensional plane (*1-8*). The wavelength-independent absorption ($\pi\alpha = 2.3\%$) as well as gate-controllable optical transition has enabled many controllable photonic devices such as photodetector (*9*), electro-optical modulator (*10*), and polarization controller (*11*). Due to the linear band structure which allows inter-band optical transitions at all photon energies, extremely large third-order optical nonlinearity $\chi^{(3)}$ has been observed in graphene (*3, 4*). As a result, graphene photonics has been extended to multifunctional nonlinear devices including mode-locked laser (*1, 2*) and optical limiter (*12, 13*). Recently, graphene transferred onto silicon photonic crystal cavity was demonstrated to be an effective nonlinear optical device enabling resonant optical switching, regenerative oscillation and four-wave mixing (*14*).

In order to assess the potential of graphene for optical signal processing, it is non-trivial to isolate the optical nonlinearities of graphene from those of the supporting substrate materials and other thermal nonlinearity effects. The intrinsic type bistable optical device is a versatile platform to investigate these nonlinearities (*15*). Significant efforts have been dedicated to develop optical bistable devices with reduced size, low operating power, short switching times and room temperature operation (*16*). A simple bistable optical device of the intrinsic type can be constructed by placing nonlinear optical materials inside a Fabry–Perot cavity (*17*). Bistability can be achieved with non-resonant nonlinearity which can induce large phase shifts and changes in the optical path length by $\lambda/2$ or more at modest laser intensity. In order to achieve large optical nonlinearity, very thick nonlinear media such as sodium vapour (*18*) and InSb thin crystal (*19*) are generally required. To date, the thinnest materials showing optical bistability are micrometer-thick GaAs superlattices (*20, 21*). The phase shift $\varphi$ of this GaAs bistable etalon can be estimated by the following standard formula:

$$\varphi = \frac{\Delta\omega}{\delta\omega}\alpha_0 d(1+\frac{\Delta\omega^2}{\delta\omega^2})^{-1} \qquad (1)$$

where $\Delta\omega$ is the laser-resonance free-frequency separation, $\delta\omega$ is the resonance half-width at half-maximum, $\alpha_0$ is the absorption coefficient and $d$ is the thickness of absorptive medium. As the change of optical phase (also optical path length) is proportional to the thickness of nonlinear media, an intriguing issue is whether this effect is observable in atomically thin, two-dimensional atomic crystals such as graphene. Although extremely large nonlinear Kerr coefficient has been reported in graphene (*3, 4*), in order to generate a phase shift approaching π, the interaction with the light field will have to be significantly enhanced beyond that allowed by an atomically thin sheet.

Here we consider the possibility of engineering vertical corrugations in the form of bubbles on graphene in order to enhance the phase shift. A graphene bubble filled with liquid can also act as an adaptive focus lens due to relative larger refractive index inside compared with air outside (*22*). Unlike planar graphene, the side wall of the curved graphene bubble provides larger interaction length with light. Considering the nonlinear Kerr effects in graphene, a graphene bubble acts as an adaptive nonlinear Kerr lens according to the finite-difference time-domain (FDTD) simulation shown in Fig 1A-E. In our simulations, a focused spot centered 643 nm below the graphene bubble (Fig. 1D) is already observable even though Kerr effect is not prominent at the power density of ~$1\times10^{10}$ W/m$^2$. This is not seen in planar graphene at the same power density (Fig. 1B). When the laser power is increased to ~$5\times10^{11}$ W/m$^2$, we find that such graphene bubble can focus light beam into a spot centered 615 nm below the mirror surface (Fig. 1E). The beam waist of the focal point is found to be reduced from ~500 nm (Fig. 1D) to ~400 nm (Fig. 1E). As a result, the phase change in graphene bubble is enhanced, which is found to be about two times larger than that of planar graphene according to our estimation (Supplementary Materials). Furthermore, the formation of graphene nanobubble uncouple the graphene film from mirror surface where the interaction with electromagnetic field could be very weak (*23*) due to its position at the resonant node of the Fabry–Perot cavity. It must be stated that the intensity-dependent self-focusing effect, a characteristic feature of adaptive Kerr lens, is found to be independent of the trapped water inside the bubble (Fig. S17 in Supplementary Materials) and is caused only by graphene.

A flat graphene sheet can be restructured to produce a high density of nanobubbles on the substrate by intense laser irradiation. Thermal stress causes the graphene to erupt into bubbles across the illuminated regions, and these bubbles are further enlarged by the gasification of trapped water molecules. Monolayer graphene grown by chemical vapor deposition (CVD) is wet-transferred onto partially reflective mirror (inset of Fig. 1F; Supplementary Materials). The quality and uniformity of the monolayer graphene film is characterized by Raman spectroscopy (Fig. 1F; fig. S1 and Supplementary Materials). The very weak D band, large 2D-to-G ratio (~ 5), and narrow G and 2D band with full-width at half-maximum (FWHM) of 22 and 38 cm$^{-1}$ respectively, indicate that the graphene used is of high crystalline quality (*24*). The graphene-covered mirror is irradiated by focused laser beam (532 nm) with a power density of ~$3.8 \times 10^{10}$ W/m$^2$ (Supplementary Materials). Subsequently, atomic force microscopy (AFM) reveals that a high density of bubbles is formed on the irradiated region. The largest graphene bubble has a diameter of ~750 nm and height of ~170 nm (Fig. 1, G and H). The side walls of the bubbles now present an added vertical dimension of ~170 nm of non-linear dispersive medium that can interact with incident radiation, which afford a much longer nonlinear dispersive path compared to a flat monolayer graphene which has a thickness of 0.335 nm. Given the fact that $\lambda/2$ is of the order of 266 nm for green laser and the finesse of the order of 17, the required index change can be calculated from $\Delta n \times 170 nm \times 17 = 266 nm$, giving $\Delta n = 0.092$. The large change in refractive index is possible on account of the large $n_2 = 1.5 \times 10^{-13}$ m$^2$/W of graphene (*3*).

Figure 2A shows our experimental setup for the observation of optical bistability in graphene nanobubbles. A high power (up to 5 W) laser at 532 nm with single longitudinal mode (TEM$_{00}$, spectral linewidth: < 5 MHz) is used as the excitation source to achieve mode-matching resonance. The Fabry-Perot interferometer with two plane mirrors is mounted on Super-Invar frame to avoid thermal expansion. The cavity is aligned to see clear and sharp interference fringes (Fig. 2B). The measured finesse of the empty cavity is ~17. In order to get a large laser power density, a lens with a focal length of 20 millimeters is used to obtain a beam waist of ~ 1 μm on the back mirror surface where the graphene sample is placed. However, the maximum transmission of the empty cavity is only of the order of 8% due to the transmission loss caused by beam walk off. At powers adequate to cause absorption saturation, we do not see any hysteresis at first indicating the ineffectiveness of a flat layer of graphene to produce a $\lambda/2$ phase shift (Supplementary Materials). However, when the cavity is subjected to higher intensities, nanobubbles form on the graphene (as verified by AFM, Supplementary Materials), leading to higher optical nonlinearity.

Optical bistable hysteresis is observed from monolayer graphene nanobubbles, as shown in Fig. 2C. In contrast, the empty cavity shows a linear response of transmitted power versus input power over the whole intensity range. A holding power (between switch-on and switch-off powers) of about 200 mW produces the widest optical hysteresis loop for the monolayer graphene over a focused beam diameter of ~1 μm. This power is comparable to that of a 5 μm thick GaAs superlattice over a focused beam diameter of 10 μm (*20*). We can see that the transmitted power increases gradually at low input power and then increases suddenly when the incident power is above 250 mW. This is because the cavity resonance is not well tuned at low intensities but shifts towards the laser frequency as the input intensity is increased. Upon lowering the input, a switch-down intensity is reached, which shifts the cavity resonance away from the laser frequency. Therefore, a hysteresis between output and input is attained (fig S8 and Movie S1 for the dynamic trace). The asymmetric shape of the hysteresis curve suggests that nonlinear response due to intensity-dependent refractive-index is the main mechanism(*18*) as absorption effects alone involve even functions of Fabry-Perot frequency mistuning and the direction of the asymmetry specifies the sign of the nonlinearity (*18, 25*).

The observed hysteretic behaviour is very much dependent on the initial cavity detuning. Figure 3A shows the bistability characteristics of the monolayer graphene nanobubbles as a function of the resonator spacing as well as cavity detuning. When the resonator spacing (i.e., mirror separation) is tuned, the cavity is driven to a mode-matched condition with the laser frequency in which the round-trip phase shift $\Phi = 4\pi n_0 L/\lambda = 2\pi N - \beta$, where $n_o$ is the linear refractive index of the medium filled in the cavity, $L$ is the cavity length, $N$ is integer and $\beta$ is the initial cavity detuning. When approaching resonance, the internal field ($I_{cavity}$) within the Fabry-Perot cavity builds up according to formula: $I_{cavity} = (1+R_B)/(1-R_B)I_{incident}T$, where $R_B$ is the reflectivity of back mirror, $I_{incident}$ is the incident intensity at the front mirror and $T$ is the transmission of the cavity. As a result, the optical nonlinearity of graphene is enhanced and we must replace $n_o$ by $n(I_{cavity}) = n_0 + n_2 I_{cavity}$, where $n_2$ is the intensity-dependent nonlinear refractive index measured in cm$^2$/W. Considering the infinitesimal thickness of graphene compared with cavity spacing (d<<L), the nonlinear phase shifts induced by graphene can be simplified as: $\Phi_{NL} = 4\pi n_2 I_{cavity} d/\lambda$ (See Supplementary Materials). At a critical value of incident intensity (near switch-on power), there is a runaway effect as the increasing $\Phi_{NL}$ increases $I_{cavity}$ and vice versa, leading to the switch-on of transmitted power from a low to a high value. Apparently the largest nonlinear phase shift is observed at resonance when the initial cavity detuning $\beta = \pi$. Upon decreasing input power, the large value of $I_{cavity}$ keeps $\Phi$ close to $2\pi N$. A sudden decrease of output power only occurs when $\Phi_{NL} - \beta = 0$ which happens at the switch-off power. As a result, the output versus input power curves illustrate different optical response ranging from limiting (at 0), differential gain (at 0.8 $\pi$), discriminator (narrow bistability at 0.83 $\pi$), to bistability (at 0.87 $\pi$, 0.93 $\pi$ and $\pi$).

To compare the transmitted signal with the reference signal, the output intensity versus time collected from photodiode I and II (Fig. 2A) are plotted in Fig. 3B. It can be seen that the transmitted power increases very slowly in the beginning, and increases dramatically once the cavity is close to the resonant threshold, thus leading to overshoot (as the maximum transmission point is not stable due to the requirement for the electric field to have a node at the mirror surface). The transmitted curves show asymmetrical shape while the incident signal shows a symmetrical ramping profile. We can see that the switch-on power is dependent on the initial cavity detuning and the widest bistable hysteresis is observed at phase $\beta=\pi$, which corresponds to $\lambda/2$ optical path length change during resonance. Compared to a recent report showing 0.2 $\pi$ phase change of terahertz wave by graphene in conjunction with metamaterial (*26*), the phase change observed here is quite large.

It is worth noting the pronounced overshoot during switch-on, which is an important characteristic of dispersive bistability. A detailed discussion on how the overshoot provides information on the optical loss is supplied in the supplementary materials. The overshoot also indicates that the device exhibits fast switching dynamics with a response time which is shorter than the overshoot time (*27*). We can estimate the cavity response by directing square waves into the cavity and monitoring the response time, as shown in Fig. 4A. It is clearly noticed that the appearance of overshoot depends on the detuning of the cavity. When the cavity approaches the mode-match condition $\beta=\pi$, it is turned on and the overshoot becomes prominent. By single exponential fitting of the decay component, the cavity is determined to have a maximum response time of around 40 ns (Fig. 4B). This switching time is much faster than thermal optical bistability effects which are typically several to 100 ms (*17*). However, the observed response time is a few orders of magnitude longer than that of the carrier relaxation time in graphene which is typically in the femtosecond range (*28, 29*). The possible explanation may have to do with the special band structure of graphene and the excited electron decay time and its dependence on the excited density of states. A pump-probe study of graphene nanobubbles at laser intensities comparable to that applied in this experiment is needed to obtain further insights.

Since the contribution from absorptive effect is negligibly small, we conclude that optical dispersive bistability, i.e., the electronic induced nonlinear refractive index change of graphene nanobubbles, is most likely the dominant mechanism for the observed bistable phenomenon (see Supplementary Materials). Indeed, our simulations based on the dispersive bistability regime manage to reproduce the experimental observations qualitatively, as shown in Fig. 4, C and D. The discrepancy in output power may originate from the fact that the cavity is not properly mode matched for a convergent beam of light, which gives rise to the low transmission of 8% even for the empty cavity. In addition, the reflection of back mirror does not go to zero at resonance due to the asymmetric reflectivity of the two mirrors used. It is also interesting to find that the optical phase shift can be further enhanced by increasing the number of layers in graphene, i.e., by transferring two layers of graphene onto the mirror and repeating the above measurements. Striking differences in terms of switching-on and holding power between monolayer graphene and bilayer are experimentally observed (Fig. 4D). The nonlinear refractive index in bilayer graphene is nearly two times that of monolayer graphene (*4*), this means only half the switch-on power (also holding power) is needed to achieve equivalent phase shift at the same initial cavity detuning. The overshoot and output power for bilayer graphene is also lower than those of monolayer graphene. This can be explained by the fact that monolayer graphene has larger transmittance (lower linear absorption) compared to bilayer graphene, which allows higher power inside the cavity according to formula $I_{cavity}=(1+R_B)/(1-R_B)I_{incident}T$, leading to a higher overshoot.

The transmission of the nonlinear cavity is about 40% that of the empty cavity, which is unexpected in view of the fact that the walls of the nanobubbles occupy less than 10% of the surface of the mirror. One explanation is that the lateral effects of the graphene walls in changing the index is not restricted to the width of the graphene layer but by the diffraction limits of the wave front which means that the entire region under a graphene bubble is being switched on at resonance. Since close to 30-40% of the surface is covered by nanobubbles (see Supplementary Materials), this may account for the high transmission. The question of whether trapped material inside the bubble contributes to the nonlinearity has been addressed by an experiment described in the supplementary materials (Fig. S12). The results show that very dry graphene film has similar nonlinear optical effect as regularly treated graphene film at different cavity detuning (Fig. S12, A and B) and their optical bistable hysteresis loops (Fig. S12C) are almost the same.

In conclusion, the exotic optical properties of graphene nanobubbles afford strong nonlinear light-matter interaction. The vertical side walls of the graphene bubbles allow a longer path length for non-linear dispersive interactions compared to planar graphene, leading to changes in the optical phase by $\pi$ and optical length by $\lambda/2$. The graphene nano-bubbles act as the adaptive Kerr lens which further enhances the phase change and drives the cavity to the resonance so as to enable the optical bistability. Graphene optical bistable devices appear to be particularly promising because of its giant optical nonlinearities and infinite small thickness, which permit the construction of miniaturized devices. They may find important applications in optical logic, memories, and analog-to-digital converters in optical signal processing systems and also be used as optical pulse discriminators and power limiters. The present study is expected to stimulate further experimental and theoretical investigations of the effects of bubbles on the non-linear properties of two-dimensional layered materials beyond graphene *(30)*.

**Acknowledgements**

K.P.L. thanks funding support from MOE Tier 2 "From in-situ observation to the growth scaling of graphene quantum dots (R-143-000-493-112)" and K. P. L. and T. V. acknowledge the NRF-SINBERISE program. Q. B. acknowledges financial support from the Lee Kuan Yew Postdoctoral Fellowship, the 863 Program (2013AA031903) and NSFC (51222208, 51290273).

**Supplementary Materials**

www.sciencemag.org/cgi/content/full/XXXX

Materials and Methods

Supplementary Text

Figs. S1 to S17

References (17)

Movie S1

**References and Notes**

1. Q. Bao *et al.*, *Adv. Funct. Mater.* **19**, 3077 (2009).
2. Z. Sun *et al.*, *ACS Nano* **4**, 803 (2010).
3. E. Hendry, P. J. Hale, J. Moger, A. K. Savchenko, S. A. Mikhailov, *Phys. Rev. Lett.* **105**, 097401 (2010).
4. H. Zhang *et al.*, *Opt. Lett.* **37**, 1856 (2012).
5. K. S. Novoselov *et al.*, *Nature* **490**, 192 (2012).
6. F. Bonaccorso, Z. Sun, T. Hasan, A. Ferrari, *Nat. Photon.* **4**, 611 (2010).
7. Q. L. Bao, K. P. Loh, *ACS Nano* **6**, 3677 (2012).
8. P. Avouris, *Nano Lett.* **10**, 4285 (2010).
9. F. Xia, T. Mueller, Y. Lin, A. Valdes-Garcia, P. Avouris, *Nat. Nanotech.* **4**, 839 (2009).
10. M. Liu *et al.*, *Nature* **474**, 64 (2011).
11. Q. Bao *et al.*, *Nat. Photon.* **5**, 411 (2011).
12. G.-K. Lim *et al.*, *Nat. Photon.* **5**, 554 (2011).
13. K. P. Loh, Q. Bao, G. Eda, M. Chhowalla, *Nat. Chem.* **2**, 1015 (2010).
14. T. Gu *et al.*, *Nat. Photon.* **6**, 554 (2012).
15. S. Smith, *Nature* **316**, 319 (1985).
16. E. Abraham, S. D. Smith, *Rep. Prog. Phys.* **45**, 815 (1982).
17. H. Gibbs, *Optical bistability: Controlling light with light*. (Academic Press, Inc., Orlando, Florida, 1985), vol. 1, pp. 481.
18. H. Gibbs, S. McCall, T. Venkatesan, *Phys. Rev. Lett.* **36**, 1135 (1976).
19. D. Miller, S. Smith, A. Johnston, *Appl. Phys. Lett.* **35**, 658 (1979).
20. H. M. Gibbs *et al.*, *Appl. Phys. Lett.* **35**, 451 (1979).
21. H. M. Gibbs *et al.*, *Appl. Phys. Lett.* **41**, 221 (1982).
22. T. Georgiou *et al.*, *Appl. Phys. Lett.* **99**, (2011).
23. A. Ferreira, N. Peres, R. Ribeiro, T. Stauber, *Phys. Rev. B* **85**, 115438 (2012).
24. A. C. Ferrari, D. M. Basko, *Nat. Nanotech.* **8**, 235 (2013).
25. T. N. C. Venkatesan, S. L. McCall, *Appl. Phys. Lett.* **30**, 282 (1977).
26. S. H. Lee *et al.*, *Nat. Mater.* **11**, 936 (2012).
27. T. N. C. Venkatesan, City University of New York (1977).

28. P. A. George *et al.*, *Nano Lett.* **8**, 4248 (2008).

29. Q. Bao *et al.*, *Nano Res.* **4**, 297 (2011).

30. Q. H. Wang *et al.*, *Nat. Nanotech.* **7**, 699 (2012)

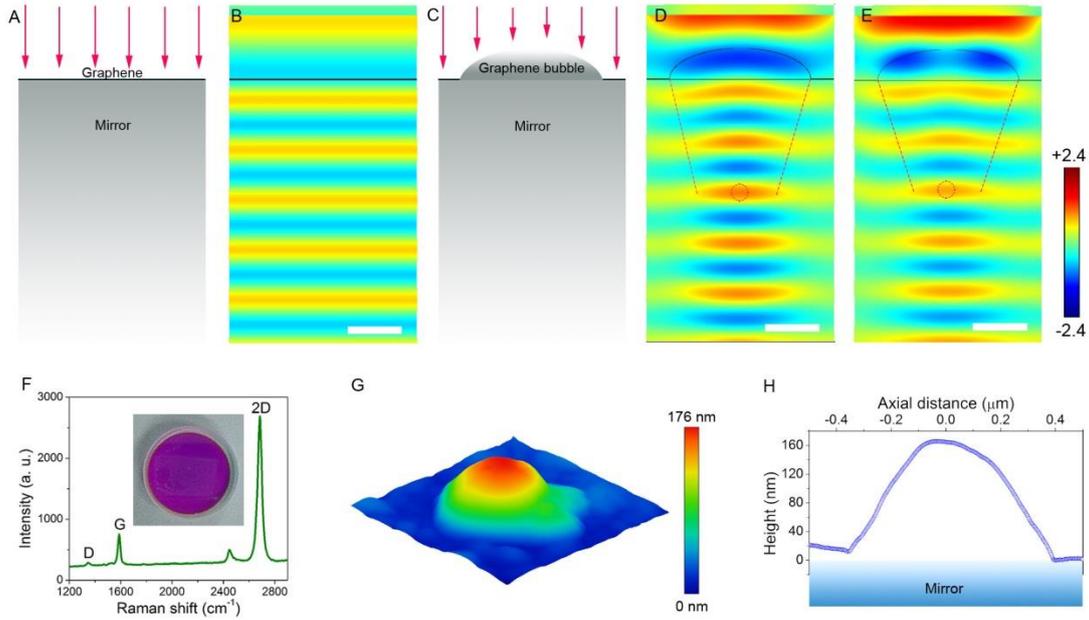

**Fig. 1**. **Graphene nanobubble and adaptive Kerr lens**. (**A**) Theoretical model of planar graphene on mirror substrate. The red arrows refer to plane light wave. (**B**) Simulated optical field of flat graphene on mirror substrate using FDTD (laser intensity: $1 \times 10^{10}$ W/m$^2$). (**C**) Theoretical model of graphene nanobubble on mirror substrate. The red arrows refer to plane light wave. (**D**) Simulated optical field of graphene nanobubble showing self-focusing effect (under a laser intensity of $1 \times 10^{10}$ W/m$^2$). (**E**) Simulated optical field on graphene nanobubble showing adaptive Kerr effect (under a laser intensity of $5 \times 10^{11}$ W/m$^2$). Scale bars in B, D and E: 300 nm. Intensity scale of local field is shown on the right. The black lines represent graphene film and the region below refers to mirror substrate. The dashed lines in red indicate the focusing effect and the dashed circles in red show the center of focal points. (**F**) Raman spectrum of monolayer graphene used in this work. Inset shows the optical image of partial reflection mirror coated with monolayer graphene. (**G**) AFM topography of graphene nanobubble formed on the mirror by irradiating with intense laser beam. Scanning area: 1×1 μm. (**H**) Cross-section histogram of the bubble obtained by AFM and schematic diagram of the bubble on mirror substrate.

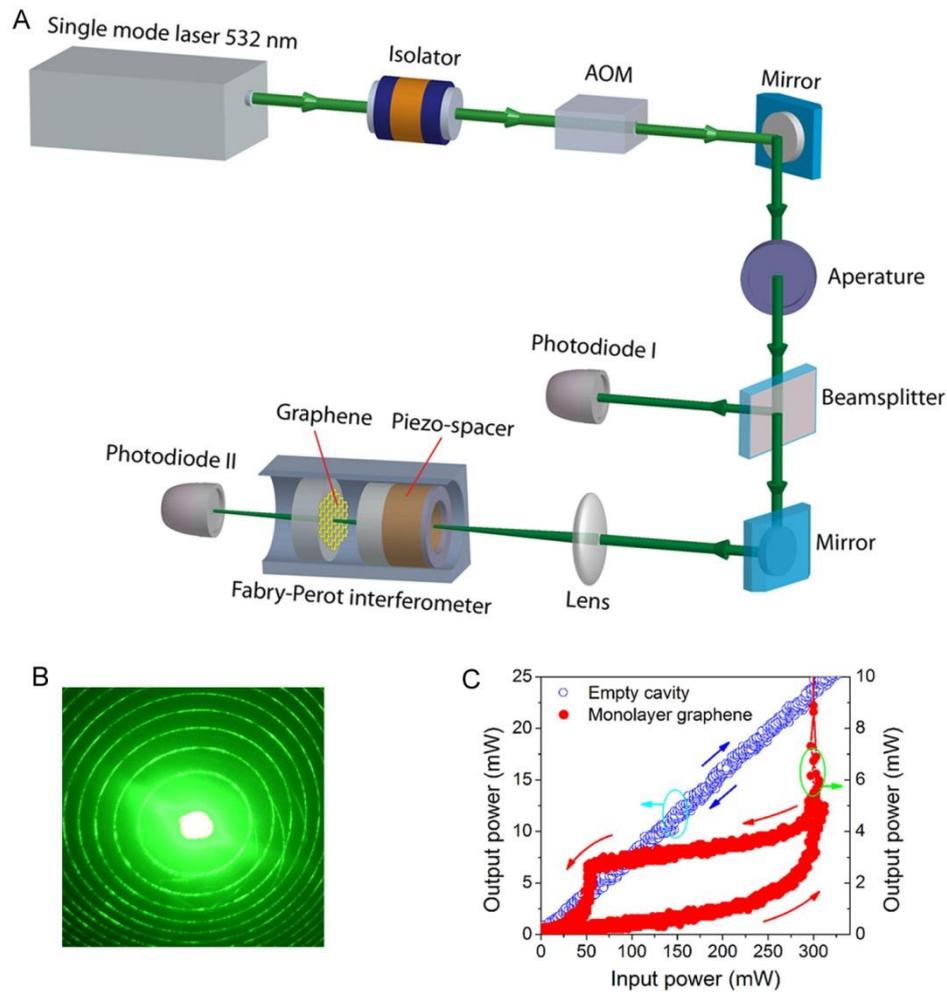

**Fig. 2. Optical bistability of graphene nanobubble.** (**A**) Experimental setup for the observation of optical bistability in graphene. (**B**) Photograph showing interference fringes from the Fabry-Perot interferometer at resonance. (**C**) Optical bistability in monolayer graphene nanobubble. The blue trace is measured from empty Fabry-Perot cavity and the red trace is obtained by coating the back mirror with monolayer graphene.

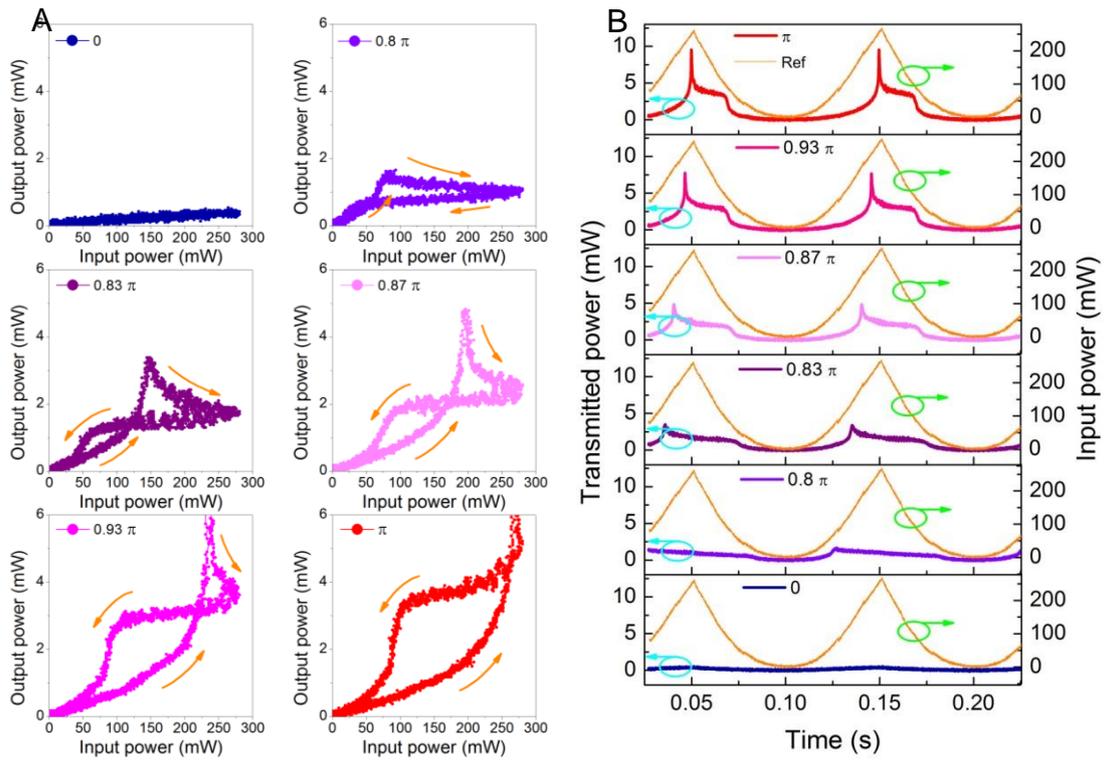

**Fig. 3. Transmission characteristic's dependence on Fabry-Perot cavity detuning.** **(A)** Optical bistable hysteresis loops as a function of resonator tuning. The cavity mistuning parameter $\beta$ was controlled by changing the offset voltage of the piezo-spacer, i.e., the cavity length was increased continuously from phase at 0 to phase at $\pi$. **(B)** Time display of transmitted signal from the Fabry-Perot cavity in comparison with reference signal (orange colour traces, right Y scale).

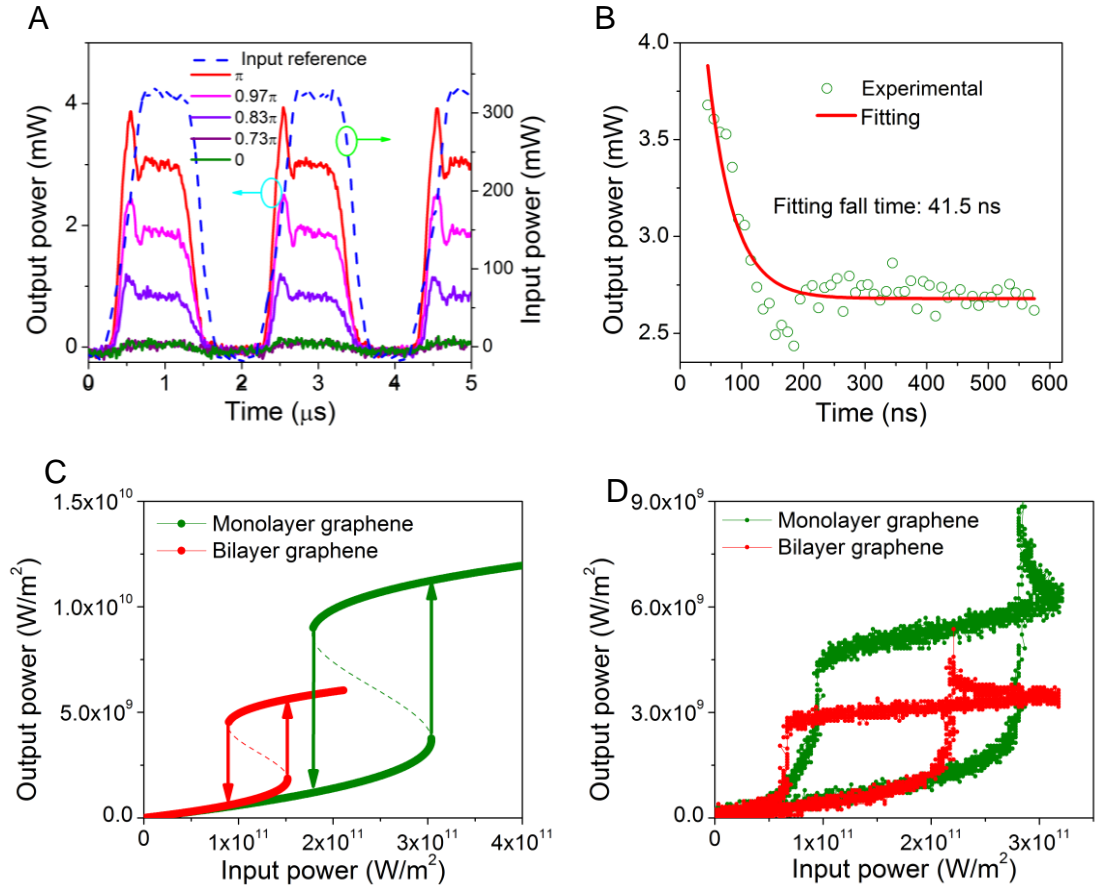

**Fig. 4**. **The dynamics of Fabry-Perot cavity and comparison of data and simulation of bistability curves.** (**A**) The input square wave (blue dash trace) and output spectra at different cavity detuning. The frequency of the square wave is 500 kHz and input power is 0.33 W. (**B**) The experimental data and the fit to the fall time of the overshoot. (**C**) Calculated optical bistability curves for monolayer and bilayer graphene. (**D**) Experimental hysteresis measured from monolayer and bilayer graphene.

# Supplementary Materials for

## Graphene Nanobubble: A New Optical Nonlinear Material

Qiaoliang Bao[1,2]*, Jianqiang Chen[3,4]*, Yuanjiang Xiang[1], Kai Zhang[1], Shaojuan Li[2], Xiaofang Jiang[1], Qing-Hua Xu[1], Kian Ping Loh[1]†, T. Venkatesan[3,4,5]†

[1]Department of Chemistry, and Graphene Research Centre, National University of Singapore, 3 Science Drive 3, 117543, Singapore.

[2]Institute of Functional Nano & Soft Materials (FUNSOM), Soochow University, Suzhou, Jiangsu 215123, China.

[3]Department of Electrical and Computer Engineering, National University of Singapore, 117576, Singapore.

[4]NUSNNI-NanoCore, National University of Singapore, 117576, Singapore.

[5]Department of Physics, National University of Singapore, 117576, Singapore.

*These authors contributed equally to this work.

†To whom correspondence should be addressed. E-mail: chmlohkp@nus.edu.sg (K. P. L.); venky@nus.edu.sg (T. V.)

**This PDF file includes:**

Materials and Methods

Supplementary Text

Figs. S1 to S17

References (17)

**Materials and Methods**

**1. Sample preparation**

**2. Characterizations of graphene bubbles**

**3. Details of the experimental setup**

**4. Response of empty cavity**

**5. Dynamic trace of optical bistability**

**6. Bistability of bilayer and multilayer graphene**

**7. Power dependent bistability**

**8. Bistability of annealed graphene**

**9. Measurements of saturation intensity**

**10. Theoretical simulations**

## 1. Sample preparation

Large area monolayer graphene films used in the work were grown by chemical vapour deposition (CVD) on Cu foil (Alfa Aesar AA13382RG). Poly(methyl methacrylate) (PMMA) thin film with thickness of ~ 100 nm was spin-coated onto the as-grown graphene film, followed by the etching of the Cu catalyst in $FeCl_3$ solution. PMMA-supported graphene films were then rinsed in deionized water thoroughly and transferred onto the surface of partial reflecting mirrors (CVI Melles Griot). The sample was then subjected to one hour baking at 60 $^o$C to remove most of the intervening moisture between graphene and mirror surface. Last, the samples were submerged into acetone (purity: HPLC, 99.9%+) for the removal of PMMA followed by a drying process with a gentle stream of $N_2$ gas. As the mirror coating is mainly a metal-oxide, the post-treatment of graphene sample in acetone will not cause any damage to the optical properties of the mirror. As-prepared samples were stored in dry cabinet before the optical measurements.

The Raman spectra and images were measured on WITEC Alpha 300 confocal micro-Raman system equipped with a 532 nm laser source and 100✕ objective lens. The representative Raman spectra from bare mirror substrate and graphene transferred onto mirror are shown in Fig. S1A. For monolayer graphene, the Raman 2D band is much stronger than G band with a 2D/G ratio of 5.0, indicating the nature of one atomic layer (*1*). The G and 2D band can be fitted well by single Lorentzian peak with full-width at half-maximum (FWHM) of 22 and 38 cm$^{-1}$ respectively (shown in Fig. S1B and S1C), which suggested that the graphene used was of high crystalline quality (*2*). These are key Raman features for monolayer graphene.

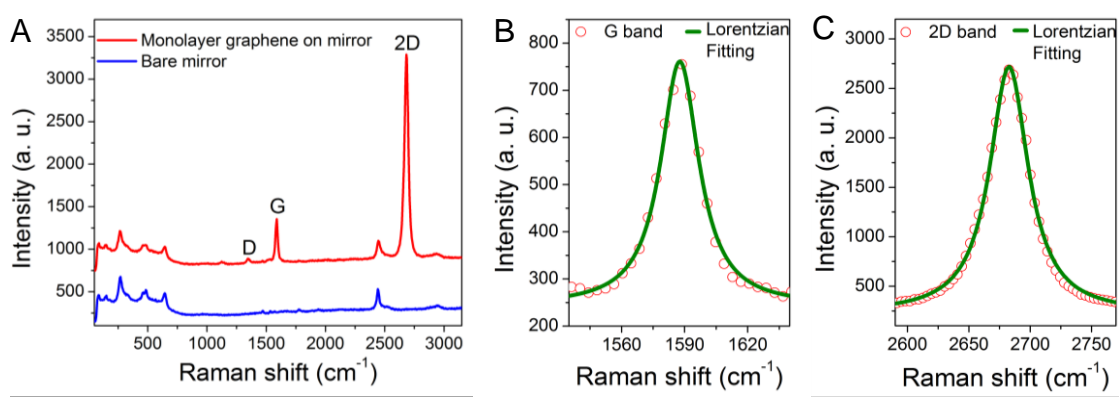

**Fig. S1**. **Raman characterization of graphene.** **(A)** Raman spectra of bare mirror substrate and monolayer graphene transferred onto mirror. **(B)** Lorentzian fitting of G band giving FWHM of 22 cm$^{-1}$. **(C)** Lorentzian fitting of 2D band giving FWHM of 38 cm$^{-1}$.

## 2. Characterizations of graphene bubbles

Based on our current techniques, it is impossible to do in-situ observation on the graphene bubbles growth under intense laser irradiation while the graphene-covered mirror is working in the Fabry-Perot cavity. It is also quite challenging to locate the grapehene bubbles on the mirror after it is removed from the Fabry-Perot cavity as the laser spot and bubbles are too small to be identified by naked eyes and will not leave any clues on the mirror surface which is observable under optical microscope. Our strategy was to duplicate the laser irradiation conditions under a micro-Raman system which is equipped with objective lens to locate the laser spot. As the Raman system is equipped with a high resolution piezoelectric stage, we were able to scan the laser in the marked area on graphene. The Raman spectrometer with time-scan and image-scan functions is able to monitor spectrum changes during the bubble growth. Atomic force microscope (AFM) is used to identify the topography of graphene bubbles in the same area which was marked and irradiated by laser.

The Raman system (WITEC Alpha 300) is equipped with a 532 nm laser with maximum output power of 50 mW. The maximum laser power reaching the mirror surface is about 30 mW, which gives a power density of ~$3.8 \times 10^{10}$ W/m$^2$ if we considered the focused beam waist to be ~1 μm (which is actually even smaller, a few hundreds nanometres, if the laser is perfectly focused on mirror surface). This power density is much higher than the saturation intensity of $1.3 \times 10^9$ W/m$^2$ in graphene, but still lower than the switching-on power in the cavity (i.e., $2.7 \times 10^{11}$ W/m$^2$).

In order to study the accumulated heating effect induced by focused laser beam, we carried out time-dependent Raman spectroscopy while setting laser power on sample surface at a modest value of 8 mW (not too high to prevent signal overload or any damage to our CCD detector). This will give a power density of ~ $1 \times 10^{10}$ W/m$^2$. Raman spectra were recorded every second (integration time: one second) for 26 seconds immediately after the laser was shift to a fresh location on graphene surface. The results are shown in Fig. S2. We found that the Raman spectrum of graphene does not change too much in the first 10 seconds, but the D band continuously increases with a strong rise of the background in the following 10 seconds. The Raman spectrum nearly remains unchanged after 20 seconds illumination. We are not very clear about the origin of the strong background. If we compare the Raman spectrum collected at 20 seconds with that collected at 1 second, we can see that the D band at 1331.2 cm$^{-1}$ is greatly enhanced, 2D band at 2668.8 cm$^{-1}$ is weakened (Fig. S2B) and G band is relatively enhanced and broadened with a peak shift of 8.1 cm$^{-1}$ from 1592.7 to 1600.8 cm$^{-1}$ (Fig. S2C). These changes in Raman spectrum agree very well with the experimental observations on graphene bubbles reported in literature (*3*), which originates from the biaxial strain(*4*).

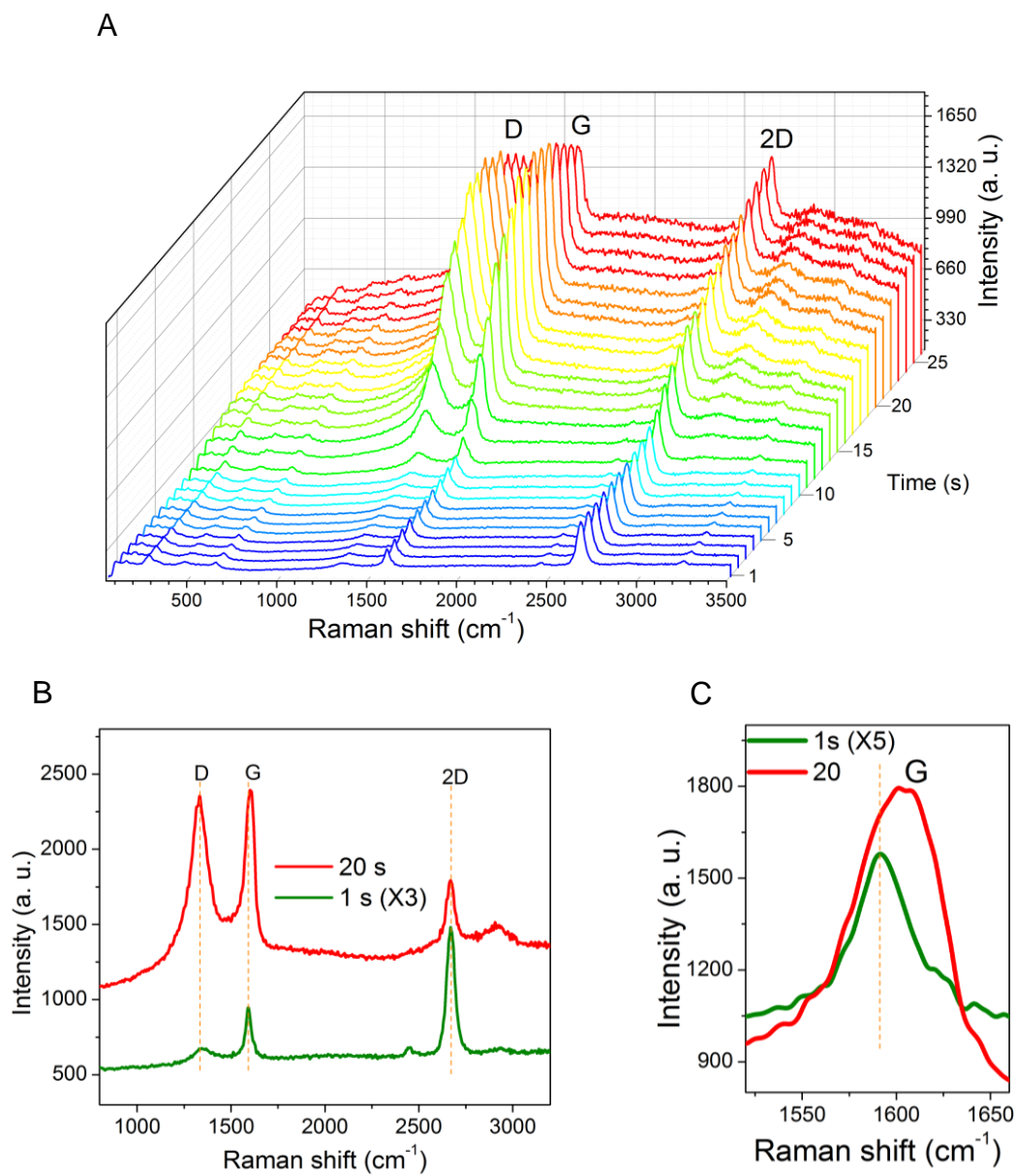

**Fig. S2. Raman characterizations of graphene nanobubbles.** (**A**) Time-dependent Raman spectra from graphene on mirror under laser irradiation. Integration time: one second. (**B**) Comparison of Raman spectra obtained at the 20th second and the first second. (**C**) Comparison of Raman G band obtained at the first second and the 20th second.

In order to verify our hypothesis on laser induced bubble formation, we carried out controlled Raman mapping and AFM measurements at the same location of the graphene sample, as shown in Fig. S3. We find a specific sample area which consists of the interface between monolayer graphene and bilayer graphene as well as a folded graphene ribbon. This feature can be easily identified in optical microscopies equipped in Raman as well as AFM system. A modest laser power (8 mW on sample surface, corresponding to $\sim 1 \times 10^{10}$ W/m$^2$) is applied to scan this area (80 points per line and a total of 80 lines) and Raman spectra were recorded simultaneously at each location. Figure S3, B and D, show the Raman images of D band, G band and 2D. As we expected, there are lots of bright spots in the Raman image of D band (indicated by arrows in Fig. S3B), corresponding to the enhancement of D band signal.

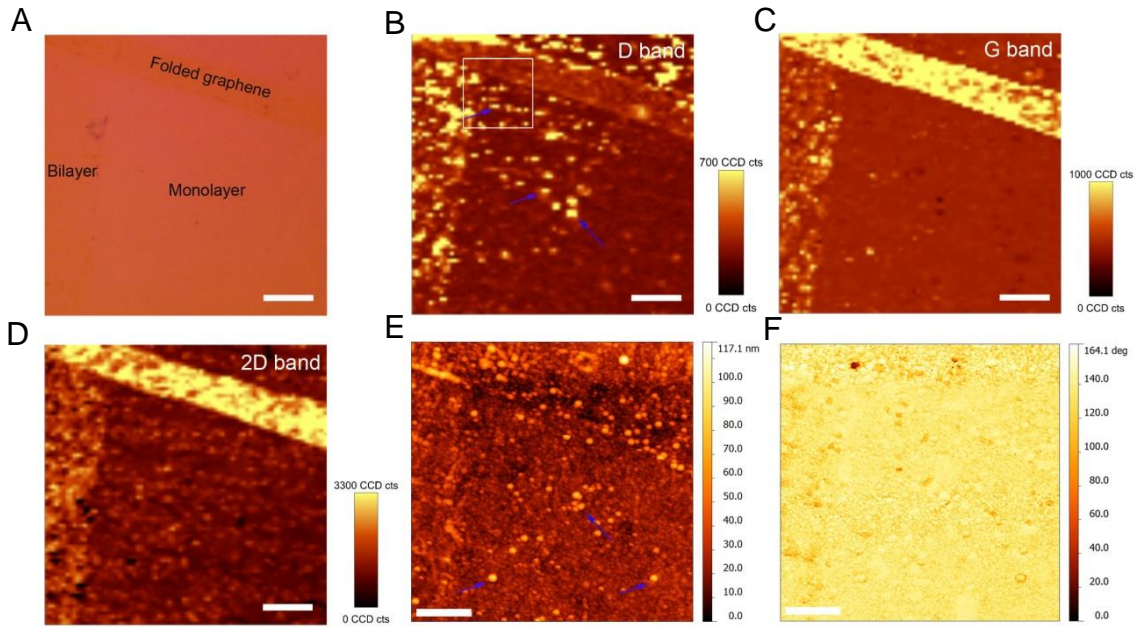

**Fig. S3. Raman and AFM characterizations on selected area of graphene film. (A)** Optical image of graphene showing the interface of monolayer and bilayer graphene as well as folded graphene. **(B-D)** Raman images of D band, G band and 2D band respectively. Scale bars: 7 μm. **(E-F)** AFM topographic and phase images of the graphene film after laser irradiation. The scanning area corresponds to the region indicated by the white square in Raman image of D band (Fig. S3B). The blue arrows in E indicate the round shape blisters. Scale bars: 2 μm.

Following AFM measurements were carried out at the upper-left corner of monolayer graphene region where we can observe strong D band signals, as shown in Fig. S3, E and F). AFM characterizations reveal that graphene bubbles can indeed be formed as thermal stress causes the graphene to erupt into bubbles across the illuminated regions. These graphene nanobubbles are found to be stable after removing the laser spot. We can see a few round shape blisters with the height larger than 50 nm, as indicated by the arrows in Fig. S3E. We can conclude that those graphene bubbles give strong D band signal due to the biaxial strain, and this is the direct evidence of graphene bubble formation after strong laser irradiation.

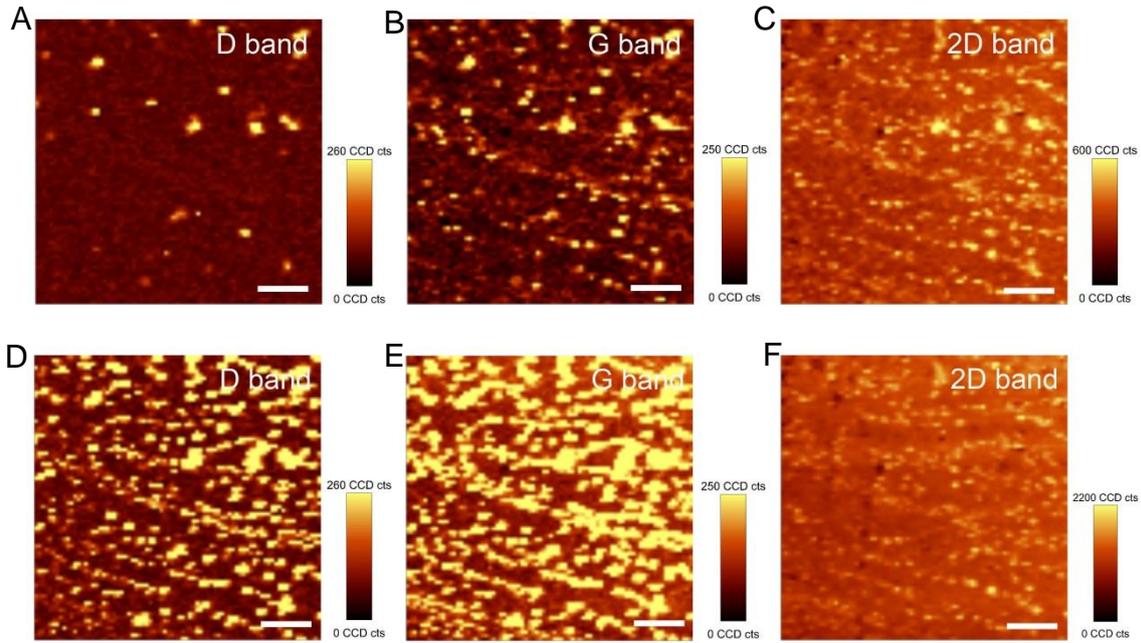

**Fig. S4. Raman images of graphene on mirror before and after laser irradiation.** **(A-C)** Raman images of D band, G band and 2D band before strong laser irradiation. **(D-F)** Raman images of D band, G band and 2D band after strong laser irradiation. Scale bars: 7 μm.

We are able to generate more spots which have similar changes in Raman signal (i.e., enhancement in D and G bands and blue-shift of G band) and locate them by Raman mapping with strong focused laser, as shown in Fig. S4. Here an area of 40×40 μm on monolayer graphene was investigated. We first performed Raman image scan in this area point-by-point (80 points per line and a total of 80 lines) with very low laser power (< 2 mW) which was below the threshold to cause any change in Raman spectra. The Raman mapping results are shown in Fig. S4, A, B and C, from which we only see a few bright spots with strong D band signals. These spots may originate from the strain of graphene film on the very rough mirror surface. Another possibility is that graphene film may be broken during the processes of transfer and removal of PMMA, which lead to defects. Subsequently, the highest laser power (30 mW on sample surface) was used to scan in the same area and the laser irradiation time at each spot was 0.3 second. At the same time, we collected the Raman spectra from each location and integrated different Raman bands to obtain the Raman images, as shown in Fig. S4, D, E and F. Obviously, we can find many bright spots which correspond to Raman signal enhancements at those specific locations in Raman images of D band and G band after strong laser irradiation. From this data, we estimate that the coverage of graphene bubbles is about 30-40% of the surface area.

In order to further investigate graphene bubble growth after strong laser irradiation ($\sim 3.8 \times 10^{10}$ W/m$^2$), we carried out more AFM measurements on graphene which has been irradiated under the strong laser beam. We were able to find the same location treated by the laser with the help of markers under the microscope of the AFM system.

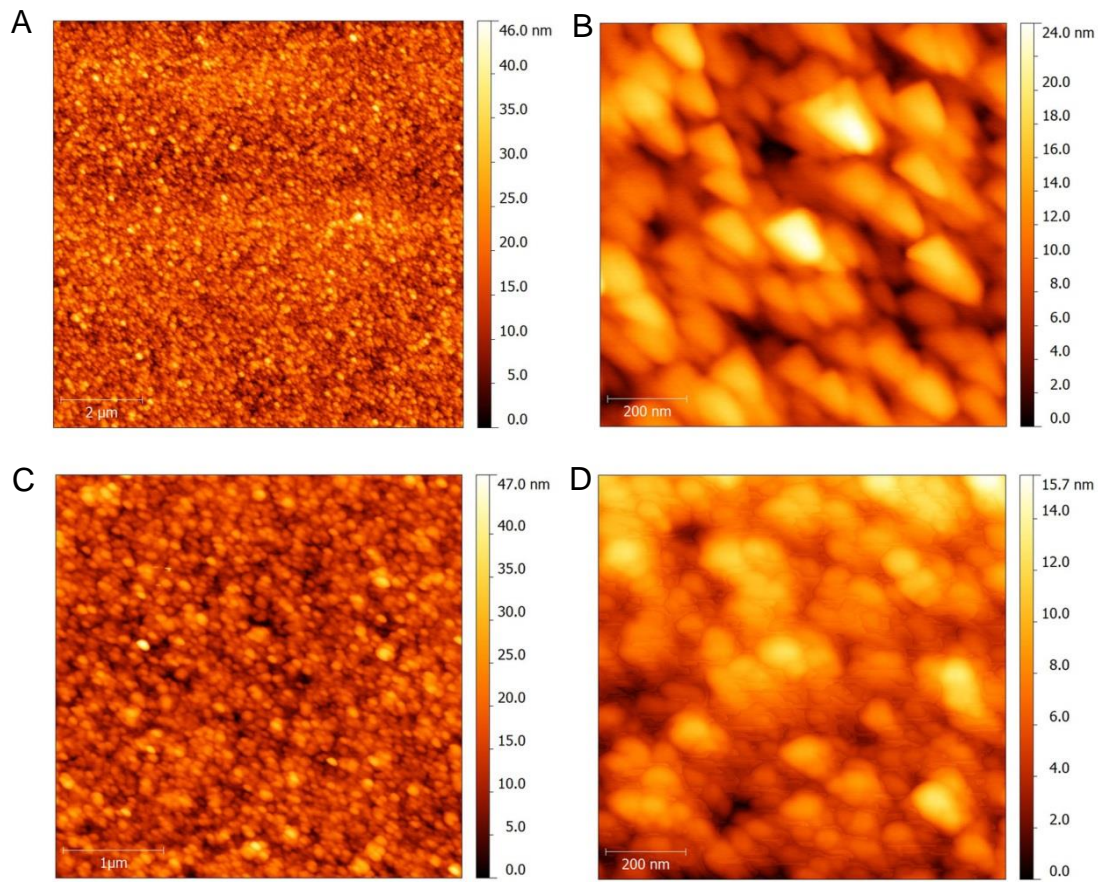

**Fig. S5. AFM characterizations of bare mirror and graphene-covered mirror.** **(A-B)** Topographic AFM images of bare mirror surface in an area of 10 μm and 1 μm respectively. **(C-D)** Topographic AFM images of graphene-covered mirror surface prior to laser irradiation in an area of 4 μm and 1 μm respectively.

Prior to laser irradiation, we did AFM measurements on bare mirror substrate and graphene-covered mirror, as shown in Fig. S5, A and B. It was found that the mirror coating consists of lots of metal oxide particles with the size less than 200 nm. As the mirror is polished to a $\lambda/10$ surface, the roughness (height difference between maximum and minimum) of the mirror is 46 nm over an area of 10 μm (shown in Fig. S5A). After covering with monolayer graphene film, the surface becomes smoother as graphene fills the gap between those particles. For example, with the same scan area of 1 μm in Fig. S5, B and D, the surface roughness has been reduced from 24 nm to 15.7 nm after covering with graphene. This is because the suspended graphene film filled the gaps among particles.

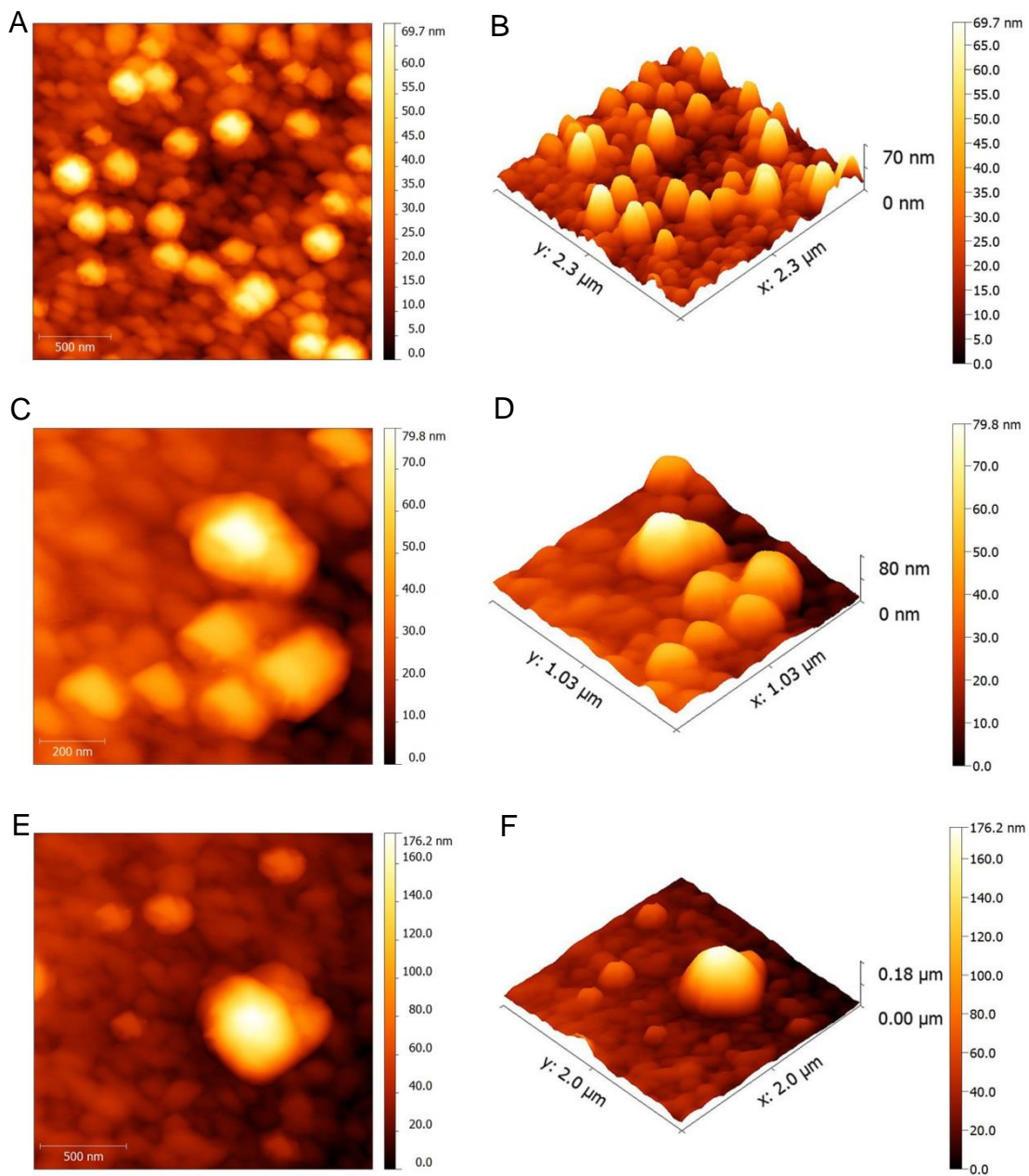

**Fig. S6. Topographic AFM images of graphene bubbles on the mirror surface after laser irradiation (~3.8×10$^{10}$ W/m$^2$). (A-B)**. 3D and 2D views of topographic AFM images showing many small graphene bubbles. **(C-D)**. 3D and 2D views of topographic AFM images showing the merging of graphene bubbles. **(E-F)**. 3D and 2D views of topographic AFM images showing single big graphene bubble.

After intense laser irradiation, AFM measurements were carried out within the same location and lots of graphene bubbles can be observed, as shown in Fig. S6. It was found that the formation of graphene bubbles depends a lot on the morphology of the mirror surface. Figure S6, A and B show a lot of small graphene bubbles with the diameter of about 200 nm and height of 50 nm. It was found that most of these bubbles are lodged on triangle shape particles which have bright contrast in the AFM topographic image. If a few particles are close to each other, the laser induced graphene bubbles may merge together to form even larger bubble, as shown in Fig. S6, C and D. Even larger graphene bubble with diameter > 500 nm and height of ~170 nm may form at certain circumstance, as shown in Fig. S6, E and F. Obviously the topographic images in Fig. S6 represent the nanobubbles at different growth stages.

Besides the heating effect, the geometry of the bubble could also be determined by the anchor points where there are strong van der Waals forces or chemical bonds between graphene and the mirror surface. The growth mechanism of these bubbles needs to be further investigated. We argue that larger bubbles can always be formed given the stronger laser irradiation or longer irradiation time even though the mirror surface is not uniform on the nanometer scale. Due to the hydrophobicity of graphene we expect any residual water to be at the mirror surface and once a stable bubble forms there will be little heat transfer to the water from the graphene and we do not expect any further vapour formation within the bubble.

### 3. Details of the experimental setup.

The detailed configuration of the experimental setup is shown schematically in Fig. 2A of maintext. A high power (up to 5 W) diode-pumped solid state laser (Newport, Millennia Edge) with single longitudinal mode (TEM$_{00}$, spectral linewidth: < 5 MHz) output at 532 nm is used in this work. A Faraday isolator is placed after the laser to prevent reflection from the Fabry-Perot cavity. The acousto-optic modulator (Isomet AOM driver, 532C-L) with a central frequency of 80 MHz and minimum rise time of 6 ns is used to modulate the input power into the Fabry-Perot cavity. The optical bistability experiments were carried out with a Burleigh RC-140 plano Fabry-Perot interferometer. Two high energy partial reflecting mirrors with surface figure of 1/10 wavelength at 532 nm are fixed in the mirror mounts of the Fabry-Perot interferometer. The front mirror has a reflectivity of 95% and the rear one has a reflectivity of 99% at 532 nm. Monolayer graphene was grown by CVD method (*5*) and wet-transferred onto the rear mirror for optical experiments. An object lens (focal length= 20 mm and beam waist= 3 mm before the cavity) is placed before the cavity to focus beam onto graphene surface with a beam waist of about 1 μm. The incident and transmitted intensities are collected by two fast silicon photodiodes with 1 ns response time and displayed as horizontal and vertical deflections on a fast oscilloscope (Tektronix, 1 GHz). The laser-induced graphene nanobubbles were characterized using a confocal Raman microscopy (WITec alpha 300) which focused the green (532 nm) laser into a small spot of about 1 μm. With the help of markers and optical microscope, the same sample area before and after intense laser illumination was characterized by AFM so as to identify the surface roughness of mirror substrate as well as geometry of graphene nanobubbles.

### 4. Response of Empty Cavity

In order to verify the response of empty cavity, we have done control experiment with the same cavity configuration but on bare mirror without graphene, as shown in Fig. S7. Obviously, the empty cavity always shows a linear response for all the phase detuning of the Fabry-Perot interferometer. This confirms that the observed bistability characteristics are due to the existence of nonlinear optical medium graphene.

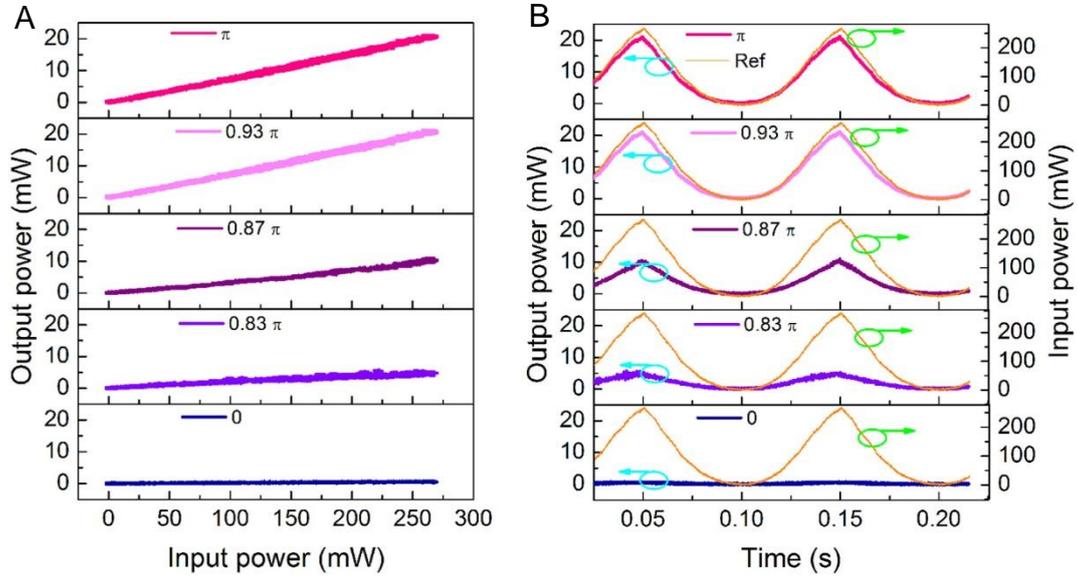

**Fig. S7. (A)** Characteristic curve dependence on Fabry-Perot cavity detuning for the empty cavity. The cavity mistuning parameter $\beta$ was controlled by changing the offset voltage of the piezo-spacer, i.e., the cavity length was increased continuously from phase at 0 to phase at $\pi$. **(B)** Time display of transmitted signal from the Fabry-Perot cavity in comparison with reference signal (orange color traces, right Y scale).

## 5. Dynamic trace of optical bistability

By setting a relatively low ramping frequency of acousto-optic modulator, the input power will change slowly so as to clearly observe the dynamic trace of the bistability hysteresis loop. Figure S8 shows the dynamic trace of the hysteresis loop of monolayer graphene in 9 seconds, in which the input power is shown on the x-axis and the output power is shown on the y-axis. As the input intensity increased, output power increased slowly at the early stage (from 1 s to 3 s), and then increased faster at the high input power range (from 4 s to 5 s) until turn on the Fabry-Perot interferometer. When the input power is decreased, the output power is firstly maintained at a high level for one second and then dropped quickly to the off state at the 7th second. Further decreasing the power (from 8 s to 9 s) will bring the interferometer back to the original state. As a result, we could resolve the dynamic trace of the hysteresis loop in one period and confirm the moving direction of the bistability.

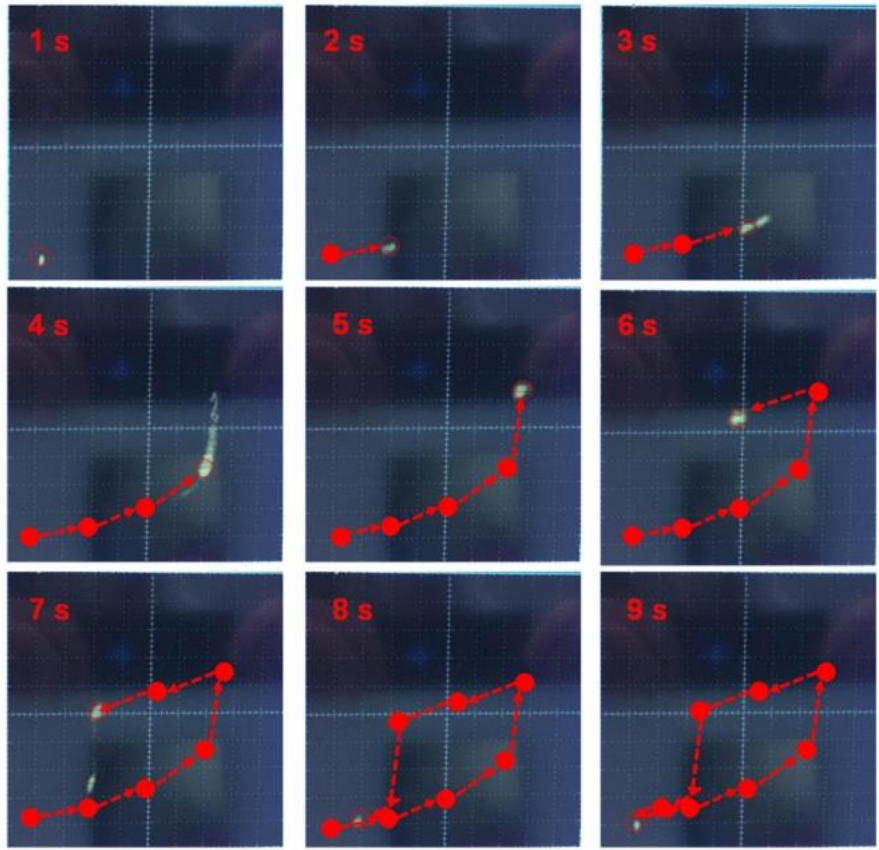

**Fig. S8. Dynamic trace of the bistability hysteresis loop from 1s to 9s by setting the ramping frequency of acousto-optic modulator as 0.1 Hz.** These pictures were captured from a video which is in a separate file.

## 6. Bistability of bilayer and multilayer graphene

In addition to the optical bistablility from monolayer graphene as discussed in the main text, we have investigated the layer-dependent bistablility from bilayer and multilayer (~10 layers) graphene, as shown in Fig. S9 and S10. Both bilayer and multilayer graphene could give clear bistability loop, and the transmitted power decreases with the increasing of graphene layers. It is found that the bistability loop from bilayer graphene is much larger than that from multilayer graphene at resonance. This is because the multilayer graphene will cause larger light absorption as well as scattering loss, which leads to the lower transmission of the cavity and furthermore limits the sharp switching between two optical states.

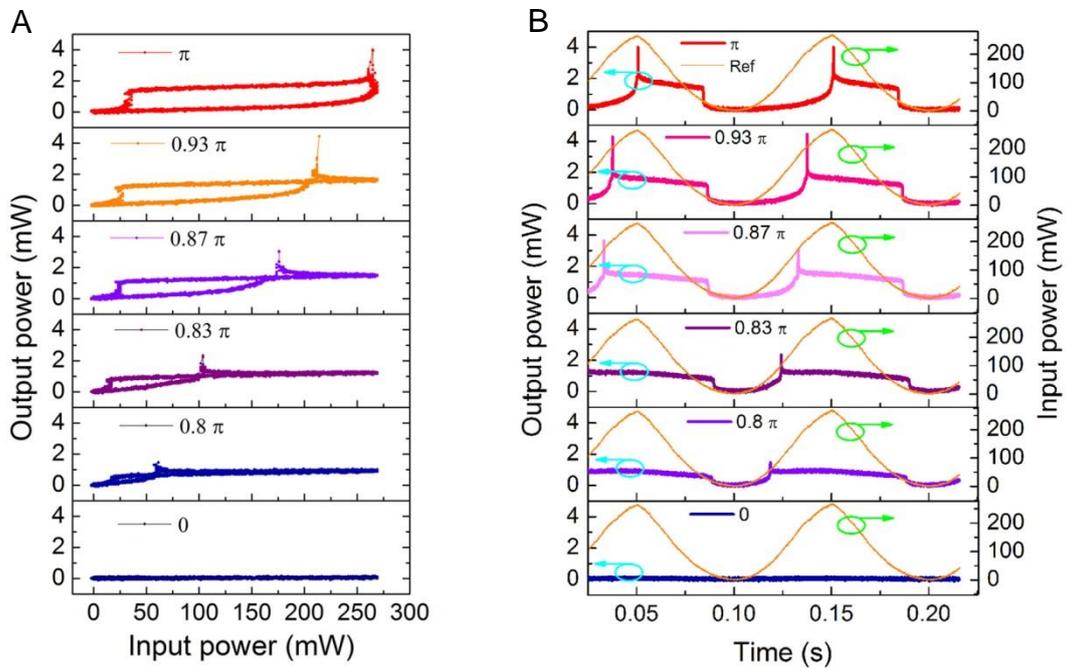

**Fig. S9. Transmission characteristic's dependence on Fabry-Perot cavity detuning for the bilayer graphene.** (**A**) Optical bistable hysteresis loops as a function of resonator tuning. The cavity mistuning parameter $\beta$ was controlled by changing the offset voltage of the piezo-spacer, i.e., the cavity length was increased

continuously from phase at 0 to phase at $\pi$. **(B)** Time display of transmitted signal from the Fabry-Perot cavity in comparison with reference signal (orange color traces, right Y scale).

It is worth noting the pronounced overshoot during switch-on, which is an important characteristic of dispersive bistability. For the monolayer (Fig. 3) and bilayer (Fig. S9) graphene nanobubbles, the maximum transmitted power at overshoot is about 40 % of transmitted power of empty cavity, which is about two times the power of switch-on state. This feature of overshoot provides clues on the optical loss (e.g., unsaturable background and beam walk-off loss) in the cavity consisting of graphene. For example, the cavity has the highest loss at $\beta=0$. When the resonator reaches the optimal condition at $\beta=\pi$, the overshoot becomes sharp and narrow, indicating the lowest optical loss in the cavity, which leads to increased trapping of the power in the cavity and results in the cavity overshoot. This overshoot appears at the high energy density range and enables the cavity to switch on because the transmission peak of the device will sweep through its maximum before settling into a lower self-consistent value (*6*).

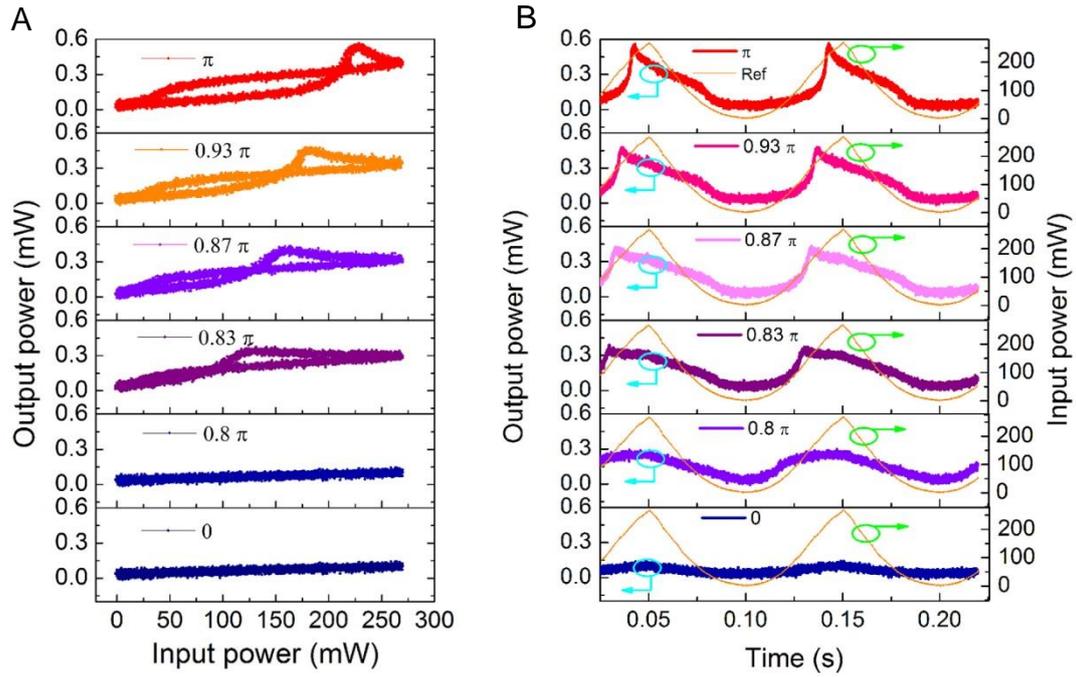

**Fig. S10. Transmission characteristic's dependence on Fabry-Perot cavity detuning for the multilayer graphene (~10 layers).** (**A**) Optical bistable hysteresis loops as a function of resonator tuning. The cavity mistuning parameter $\beta$ was controlled by changing the offset voltage of the piezo-spacer, i.e., the cavity length was increased continuously from phase at 0 to phase at $\pi$. (**B**) Time display of transmitted signal from the Fabry-Perot cavity in comparison with reference signal (orange color traces, right Y scale).

## 7. Power dependent bistability

Figure S11 shows the power-dependent bistability of the monolayer, bilayer and multilayer (~10 layers) graphene. As discussed earlier in the main text, cavity turn-on power could be controlled by tuning the cavity length. We first put the cavity at the best alignment condition as well as suitable cavity detuning for resonance at lower laser power of 1 W, then gradually increase the laser power and ramp the acousto-optic modulator and record the bistability traces. We can only observe that the overshoot tails are prolonged with the increase of incident laser power. It is interesting to note that turn-on and turn-off power were almost independent on the incident light power. This is correct for monolayer and bilayer graphene, attesting to the non-thermal bistable nature of our device (*7*).

However, the bistability hysteresis loop is not so obvious at lower input power for multilayer graphene. The low transmission of the multilayer graphene may result from the increased absorption of the incident light by the increased thickness of the graphene. The scattering loss originated from the rough surface of multilayer graphene may also lead to the low transmission and imperfect bistability. Nevertheless, bistability hysteresis loop could be operated at relatively low input power, which indicates possible applications for optical switching and computing.

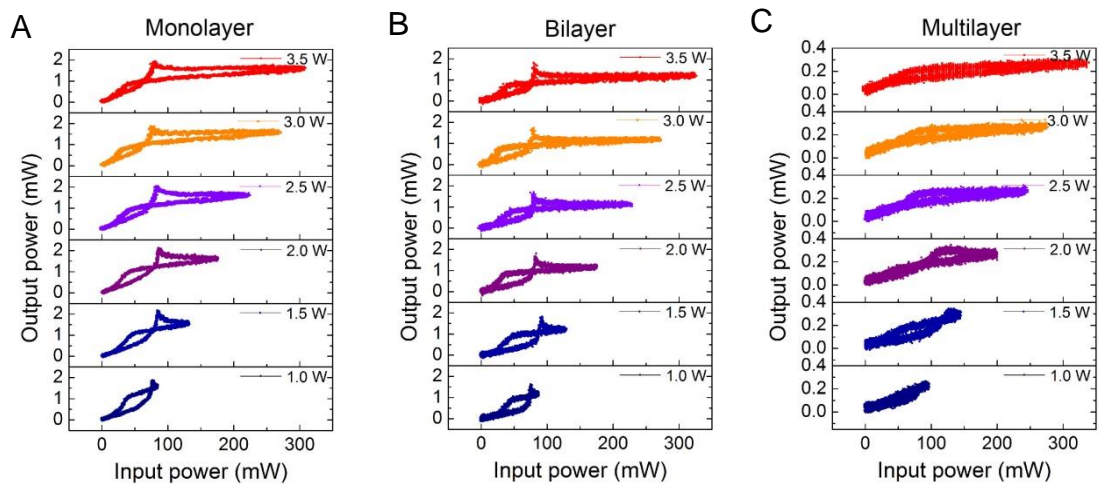

**Fig. S11. Power dependent bistability of the monolayer, bilayer and multilayer (~10 layers) graphene.** The laser power is tuned from 1 W to 3.5 W. The input power at X-axis represents the real incident power which is directed into Fabry-Perot cavity.

## 8. Bistability of annealed graphene

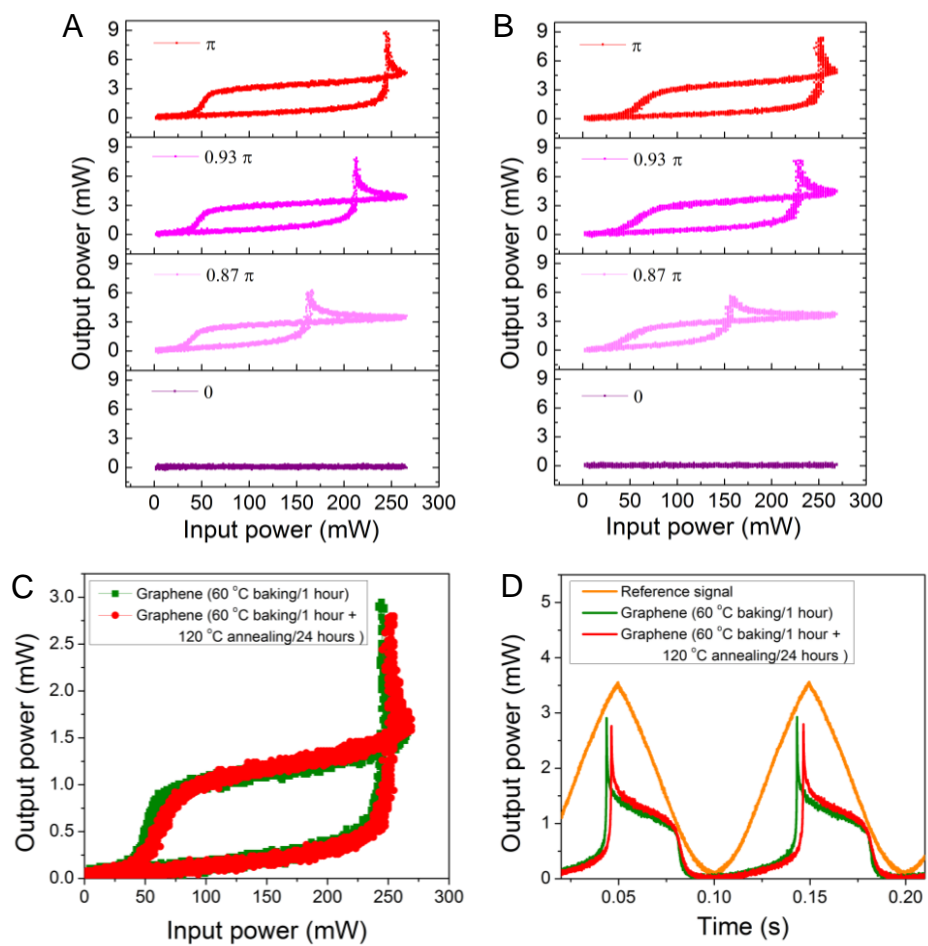

**Fig. S12. Optical bistability of non-annealed and annealed monolayer graphene.** **(A)** Optical bistable hysteresis loops of non-annealed monolayer graphene (baked only at 60 °C for one hour before removing PMMA). **(B)** Optical bistable hysteresis loops of annealed monolayer graphene (baked at 60 °C for one hour and annealed at 120 °C for 24 hours before removing PMMA). **(C)** Comparison of optical bistable hysteresis loops of non-annealed and annealed monolayer graphene. **(D)** Time display of transmitted signal from the Fabry-Perot cavity with non-annealed and annealed monolayer graphene.

In order to investigate whether the material (i.e., air, moisture and other residues) locked in the nanobubble contribute to the optical nonlinear effect, we carried out control experiments on the very dry graphene film and compare its optical bistability with that of regularly treated graphene film. The strategy is to anneal graphene samples at elevated temperature for long time to expel most of the intervening moisture and other residues prior to any optical measurements. In this control experiment, graphene-covered mirror was annealed at 120 $^{o}$C for 24 hours after graphene transfer, leading to a very dry graphene film. A reference graphene sample was also prepared by a regular wet-transfer technique (as introduced in Section 1) for comparison.

Fig. S12 shows the optical bistability results of non-annealed and annealed monolayer graphene. The non-annealed graphene sample was just baked at 60 $^{o}$C for one hour before removing PMMA. The annealed graphene sample was baked at 60 $^{o}$C for one hour and heated at 120 $^{o}$C for 24 hours before removing PMMA. After removing PMMA, the sample was further heated at 120 $^{o}$C for 8 hours just before the optical measurements. It was found that the very dry graphene film has similar nonlinear optical effect as regularly treated graphene film at different cavity detuning (Fig. S12, A and B) and their optical bistable hysteresis loops (Fig. S12C) are almost the same. The small differences at the switch-on and switch-off points are most likely due to the slightly different cavity spacing and alignment in these two separate experiments. These results support the argument that the material locked in the graphene nanobubbles will not cause significant optical effect and the bistability observed in our experiments is only due to the nonlinearity of the graphene.

## 9. Measurements of saturation intensity

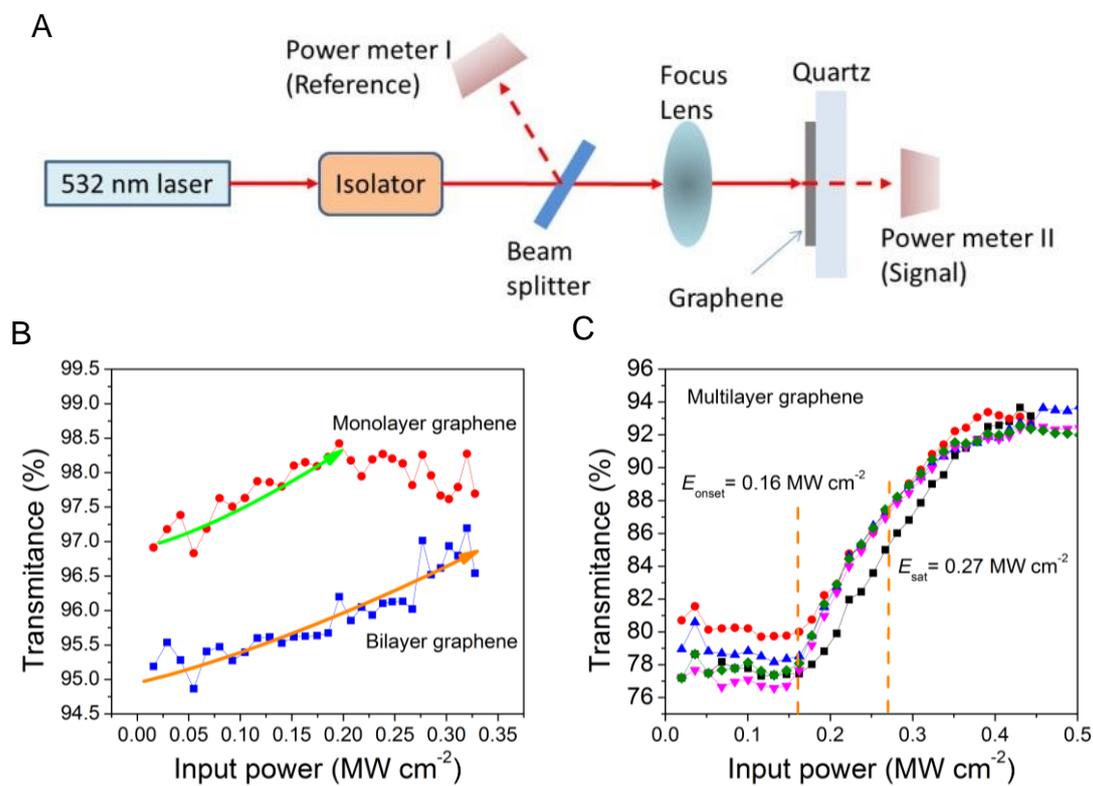

**Fig. S13**. Saturation intensity measurements on graphene films. **(A)** Schematic diagram of the experimental setup. The arrows are just guides for eyes. **(B)** Saturable absorption of monolayer and bilayer graphene. **(C)** Saturable absorption of multilayer graphene (~ 10 layers). The traces represent the measurements at different positions where the thickness is a little bit different.

In order to verify the operating mechanism of the optical bistability, it is non-trivial to find out the saturation intensity in graphene samples which we have used. We transferred the same batch of monolayer and bilayer (stack of two monolayers) as well as multilayer graphene onto quartz substrate for the measurements. The experimental setup is shown in Fig. S13A. By using a focus lens, the laser spot is confined to an area of about 1 μm. We are able to gradually increase the laser power and record the transmission of the graphene film at each input power. The nonlinear transmittance results are shown in Fig. S13, B and C. For monolayer and bilayer graphene, the initial transmittance at low input power starts from ~97% and ~95% respectively. While increasing the input power of our 532 nm laser, the transmittance increases and then reaches saturation, which is a typical behavior of saturable absorption. This saturable absorption phenomenon is even obvious in multilayer graphene which has a linear transmittance of about 78% (corresponding to ~10 layers of graphene). Saturation intensity is defined as the optical intensity required in a steady state to reduce the absorption to half of its unbleached value.(*8*) The saturation intensity is estimated to be ~0.13 MW/cm$^2$ ($1.3 \times 10^9$ W/m$^2$) for monolayer graphene, ~0.18 MW/cm$^2$ ($1.8 \times 10^9$ W/m$^2$) for bilayer graphene and ~0.27 MW/cm$^2$ ($2.7 \times 10^9$ W/m$^2$) for multilayer graphene (~10 layers).

## 10. Theoretical simulations
## 10.1. The phase in the Fabry-Perot cavity

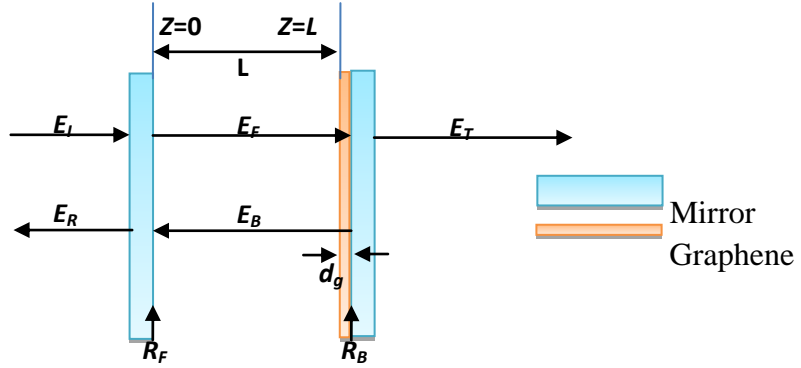

**Fig. S14.** Schematic model of the Fabry-Perot interferometer. $E_I$, $E_R$, $E_B$ and $E_T$ are the incident, reflected, forward, backward and transmitted electric fields, respectively. $R_F$ and $R_B$ refer to the reflectivity of the front and back mirror. $d_g$ refers to the thickness of graphene film and $L$ refers to the cavity length.

The Fabry-Perot cavity containing graphene sheets is considered in free space, as shown in Fig. S14. In the free space, the light is freely propagating and the phase change of one-way transmission can be written as,

$$\varphi_0 = \frac{2\pi}{\lambda} n_0 L, \tag{1}$$

where $L$ is the free-space propagating distance, $n_0$ is the refractive-index of the vacuum, and $\lambda$ is the operating wavelength of the laser.

The graphene sheet is coated on the back mirror of the Fabry-Perot cavity. It is well known that the transmittance of monolayer graphene is $T = 1 - \alpha\pi = 97.7\%$, and the transmittance is of few-layer graphene can be still maintain at >90%. Here, for the sake of simplicity, we assume that the transmittance of graphene (monolayer and few-layer) is 1. Therefore, the interfacial effects between the free space and graphene sheet can be neglected. The influence of the graphene sheet on the light propagating in the Fabry-Perot cavity mainly reflects on the phase change. If we assume that the thickness of the graphene layers is $d_g$, the phase change induced by the graphene

sheet is

$$\varphi_1 = \frac{2\pi}{\lambda} n_g d_g, \qquad (2)$$

where $n_g$ is the effective refractive-index of the graphene sheets.

Considering the nonlinear optical effect in graphene, the effective refractive-index can be written as

$$n_g = n_g' + \Delta n_g, \qquad (3)$$

where $n_g'$ is the linear refractive-index, $\Delta n_g$ is the refractive index changes caused by the nonlinear effect. Therefore, the phase change induced by graphene can be rewritten as

$$\varphi_1 = \frac{2\pi}{\lambda}\left(n_g' + \Delta n_g\right)d_g = \varphi_1' + \Delta\varphi(I), \qquad (4)$$

where $\varphi_1' = \frac{2\pi}{\lambda} n_g' d_g$ is the phase change due to the linear refractive-index, $\Delta\varphi(I) = \frac{2\pi}{\lambda} d_g \Delta n_g$ is the phase change due to the nonlinear optical effect depending on the light intensity in the Fabry-Perot cavity.

Thus, the total phase change of the Fabry-Perot cavity can be written as

$$\Phi = \left(\frac{2\pi}{\lambda} n_0 L + \frac{2\pi}{\lambda} n_g' d_g\right) + \frac{2\pi}{\lambda} d_g \Delta n_g. \qquad (5)$$

Considering Kerr nonlinear effect in graphene, the nonlinear refractive index change is expressed as

$$\Delta n_g = n_2 I_{\text{cavity}} \approx n_2 I_{\text{eff}}, \qquad (6)$$

where $n_2$ is the nonlinear refractive index of graphene, $I_{\text{cavity}}$ is the laser intracavity intensity which can be approximately treated as the average light intensity ($I_{\text{eff}}$) in the cavity.

The relation between average light intensity $I_{\text{eff}}$ in the cavity and the transmission light intensity $I_T$ is (9, 10)

$$I_{\text{eff}} = \frac{1+R_B}{1-R_B} I_T, \qquad (7)$$

where $R_B$ is the intensity reflectivity of the back mirror. Hence, the total phase of the Fabry-Perot cavity is

$$\Phi = \left(\frac{2\pi}{\lambda}n_0 L + \frac{2\pi}{\lambda}n_g' d_g\right) + \frac{2\pi}{\lambda}d_g n_2 \frac{1+R_B}{1-R_B}I_T = \Phi_0 + \beta_2 I_T \tag{8}$$

where $\Phi_0 = \frac{2\pi}{\lambda}(n_0 L + n_g' d_g)$ is the phase due to the linear effects, $\beta_2 = \frac{2\pi}{\lambda}d_{eff} n_2$, $\beta_2 I_T$ is the phase change caused by the nonlinear effects, and $d_{eff} = d_g \frac{1+R_B}{1-R_B}$.

## 10.2. Dispersive and absorptive optical bistability in Fabry-Perot cavity

**(1) Dispersive optical bistability**

We consider the situation illustrated in Fig. S14 of a Fabry-Perot cavity with back mirror coated by a graphene and driven by an injected field $E_I$. Here, for the sake of simplicity, we also assume that the transmittance of the graphene sheets are 1, hence the interfacial effects between the free space and graphene sheets can be neglected. The influence of the graphene sheets on the light propagating in the Fabry-Perot cavity mainly manifests as the phase shift. In order to determine the field inside the Fabry-Perot cavity, we write the boundary conditions as,

$$E_T = \sqrt{T} E_F(L + d_g) \tag{9}$$

$$E_F(0) = \sqrt{T} E_I + R\, e^{i2\varphi} E_F(0), \tag{10}$$

where $\varphi$ is the phase giving by Eq. (8), $E_F(0)$ and $E_F(L+d_g)$ are the electric field at $Z=0$ and $Z=L+d_g$, respectively. Here, we assume that front and back mirrors have the same reflectance, i.e., $R_F = R_B$. $E_I$, $E_R$, $E_F$, $E_B$, $E_T$ are the incident, reflected, forward, backward, and transmitted electric field slowly varying complex amplitudes, respectively. If we consider the linear absorption of the graphene sheets, then the phase $\varphi$ can be written as,

$$\Phi = \Phi_0 + \beta_2 I_T + i\alpha d_g/2, \tag{11}$$

$\alpha$ is the absorption coefficient. As such, we assume that the absorption coefficient

depends only on the "uniform field" approximation (field envelope is position independent) and thereby neglect saturation or nonlinear index corrections due to field changes along the laser axis.

Based on Eq. (9-11), we have

$$\frac{E_F(0)}{E_I} = \frac{\sqrt{T}}{1 - R\, e^{2i(\Phi_0 + \beta_2 I_T)} e^{-\alpha d_g}} \tag{12}$$

The field $E_T$ is simply related to $E_F$,

$$E_T = \sqrt{T} E_F(L + d_g) = \sqrt{T} e^{-\alpha d_g} e^{2i(\Phi_0 + \beta_2 I_T)} E_F(0) \tag{13}$$

Combining this with Eq. (12), we have the amplitude transmission function

$$\frac{E_T}{E_I} = \frac{T e^{-\alpha d_g} e^{2i(\Phi_0 + \beta_2 I_T)}}{1 - R\, e^{2i(\Phi_0 + \beta_2 I_T)} e^{-\alpha d_g}}. \tag{14}$$

As an arbitrary complex absorption coefficient ($\alpha$), our equations can be used to investigate both purely dispersive and purely absorptive optical bistability.

Now, we consider the purely dispersive case [$\text{Re}(\alpha) = 0$]. This case is obviously an approximation, since a nonlinear dispersion implies the existence of a nonlinear absorption, but the latter decreases significantly faster as the laser frequency is detuned from the medium's resonances. Setting

$$\beta = \text{nearest multiple of } (2\pi) - 2(\Phi_0 + \beta_2 I_T) \tag{15}$$

$\beta$ is the cavity-laser phase detuning, we find that the amplitude transmission function in Eq.14 yields

$$\frac{E_T}{E_I} = \frac{T}{e^{-2i(\Phi_0 + \beta_2 I_T)} - R}. \tag{16}$$

The intensity transmission function is

$$\frac{I_T}{I_I} = \frac{T^2}{|e^{i\beta} - R|^2} = \frac{1}{1 + 4R \sin^2(\beta/2)/T^2} \tag{17}$$

where $I_T = |E_T|^2$, $I_I = |E_I|^2$ (here we suppose the electric fields (e.g., $E_T$, $E_I$) are dimensionless fields corresponding to the usual dimensionless intensity definition). Combining Eq. (15) and (17), we are able to calculate the hysteresis trace of optical bistability.

**(2) Dispersive optical bistability with linear absorption**

If we consider the graphene sheets with linear absorption and different intensity reflectivity for the front and back mirror, then the total Fabry-Perot intensity transmission is(10)

$$T = \frac{(1-R_B)(1-R_F)(1-A)}{(1-R_\alpha)^2} \frac{1}{1+F\sin^2(\beta/2)} \quad (18)$$

Here, $A = 1 - e^{-\alpha_0 d_g}$ is the intensity absorption per pass, $\alpha_0$ is the linear absorption coefficient, $R_F$ and $R_B$ are the intensity reflectivity of the front and back mirror, respectively. $R_\alpha = (1-A)\sqrt{R_F R_B}$ is the effective mean reflectivity, $F = 4R_\alpha/(1-R_\alpha)^2$, the fitness of Fabry-Perot cavity is $\mathbb{F} = (\pi/2)\sqrt{F}$, and $\beta/2$ is the round-trip phase, it has the same value as that in Eq (15).

Where $I_{eff}$ can be explicitly written as

$$I_{eff} = I_I \frac{A}{\alpha_0 d_g} \frac{(1-R_F)(1-R_B)}{(1-R_\alpha)^2} \frac{1}{1+F\sin^2(\beta/2)} \quad (19)$$

$I_I$ is the incident intensity. The combination of Eq. (18) and Eq. (19) gives another expression (parametric in $I_{eff}$) of the intensity transmission $T$ in Fabry-Perot cavity

$$T = \frac{\alpha_0 d_g}{A} \frac{(1-R_B)(1-A)}{(1+R_{B\alpha})} \frac{I_{eff}}{I_I} \quad (20)$$

where $R_{B\alpha} = (1-A)R_B$ is the effective reflectivity of the back mirror.

Equations (18) and (20), which can be solved simultaneously to eliminate $I_{eff}$, describe the nonlinear Fabry-Perot action with linear absorption, and the behavior of the nonlinear cavity can be similarly visualized by graphic solution of Eq. (18) and Eq. (20).

**(3) Absorptive optical bistability**

The first theoretical model of purely absorptive optical bistability was proposed by Szoke et al.(*11*) in 1969. Based on the boundary conditions for the light interaction in the Fabry-Perot cavity as well as the saturation equations of a homogeneously broadened two-level system, the condition purely absorptive bistability is $\alpha d/(T+\alpha_B d)>8$, where $\alpha_B$ is unsaturable background absorption coefficient, $d$ is the thickness of absorptive medium, and $T$ is transmittance. Considering the planar monolayer graphene has a thickness of 0.335 nm and an absorption coefficient of 301,655 $cm^{-1}$ (*12*), we arrive at $\alpha d=0.01$, which is too small to fulfil the requirement for purely absorptive bistability. Even if we consider the vertical dimension of largest graphene bubble which gives ~170 nm light path length, we arrive at $\alpha d=5.13$. This means it will be fruitless to search for purely absorptive bistability in our case.

In order to further verify above assumptions, we calculate the intensity transmission trace, i.e., output intensity as a function of input intensity, based on the experimental parameters of monolayer graphene. In the case of purely absorptive optical bistability, the input field frequency coincides with both a cavity resonance and the atomic line center. Ignoring the unimportant phase factor in the numerator of Eq. (14), and supposing that $\alpha d_g \ll 1$ so that $e^{-\alpha d_g} \approx 1-\alpha d_g$, Eq. (14) can be simplified as

$$\frac{E_T}{E_I}=\frac{T}{1+\alpha d_g - R}=\frac{1}{1+\alpha d_g/T} \tag{21}$$

If $\alpha d_g/T$ is larger, i.e., $T \ll \alpha d_g \ll 1$, then $E_T/E_I$ is small. However, if the absorption can be bleached, $E_T/E_I$ can approach unity transmission.

Specifically for a two-level atom, it shows that on resonance $\alpha=\alpha_s/(1+I)$, where $I$ is given in units of the saturation intensity $I_s$, and $\alpha_s$ is the saturable absorption coefficient. For convenience, we take $E_I$ and $E_T$ also in the corresponding amplitude units, which gives $I=I_T/T$. Combining these formulas

with Eq. (18) and solving for $E_I$, we find

$$\frac{E_I}{\sqrt{T}} = \frac{E_T}{\sqrt{T}}\left[1 + \frac{\alpha_s d_g / T}{1 + I_T / T}\right] \quad (22)$$

The intensity transmission function is

$$\frac{I_I}{T} = \frac{I_T}{T}\left[1 + \frac{\alpha_s d_g / T}{1 + I_T / T}\right]^2 \quad (23)$$

The simulated transmittance curves with purely absorptive effect are shown in Fig. S15. For the both cases of planar monolayer graphene ($\alpha d = 0.01$) and bubble of monolayer graphene ($\alpha d = 5.13$), we are not able to reproduce the bistable hysteresis loop in the experiments.

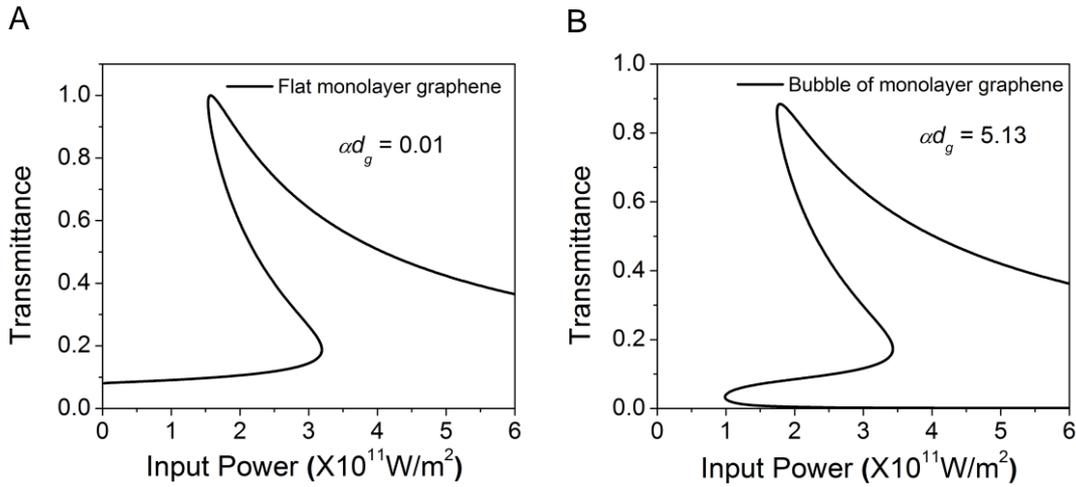

**Fig. S15. Simulated transmittance traces based on the model of purely absorptive bistability.** (**A**) Planar monolayer graphene. (**B**) Bubble of monolayer graphene with ~170 nm light path length.

**(4) Combination of absorptive and dispersive optical bistability**

Equation (16) can written as,

$$\frac{I_T}{I_I} = \frac{T^2}{\left|e^{\alpha d_g + i\beta} - R\right|^2} \quad (24)$$

For the materials with both linear absorption and nonlinear absorption

$\alpha = \alpha_0 + \alpha_{NL} = \alpha_0 + \dfrac{\alpha_s}{1+I/I_s}$, Eq. (24) can be written as,

$$\dfrac{I_T}{I_I} = \dfrac{T^2}{\left|e^{(\alpha_0+\alpha_{NL})d_g + i\beta} - R\right|^2} = \dfrac{T^2 e^{-2(\alpha_0+\alpha_{NL})d_g}}{1 - 2R\cos\beta e^{-(\alpha_0+\alpha_{NL})d_g} + R^2 e^{-2(\alpha_0+\alpha_{NL})d_g}} \quad (25)$$

or

$$\dfrac{I_I}{I_T} = \dfrac{1 - 2R\cos\beta e^{-(\alpha_0+\alpha_{NL})d_g} + R^2 e^{-2(\alpha_0+\alpha_{NL})d_g}}{T^2 e^{-2(\alpha_0+\alpha_{NL})d_g}} \quad (26)$$

Using the relation $I = I_T/T$, we can simulate the optical bistability curve by using the following equation,

$$\dfrac{I_I}{I_T} = \dfrac{1 - 2R e^{-\alpha_0 d_g}\cos\left(\beta_0' + \beta_2 I_T\right) e^{-\frac{\alpha_s d_g}{1+I_T/(TI_s)}} + R^2 e^{-2\alpha_B d_g} e^{-\frac{2\alpha_s d_g}{1+I_T/(TI_s)}}}{T^2 e^{-2\alpha_0 d_g} e^{-\frac{2\alpha_s d_g}{1+I_T/(TI_s)}}} \quad (27)$$

where $\beta_0' = \Phi_0 - 2m\pi$.

**(5) The nonlinear phase shifts of the planar graphene and graphene bubble**

In the Fabry-Perot cavity, the forward light field is assumed to be a plane wave with Gaussian distribution,

$$E(x,y,z) = E_0 \exp\left(-\dfrac{x^2+y^2}{w^2}\right)\exp(-jkz) \quad (28)$$

Moreover, it is supposed that the light field distribution in space does not change after reflection by the back mirror of the cavity, then the reflected light is,

$$E_r(x,y,z) = rE_0 \exp\left(-\dfrac{x^2+y^2}{w^2}\right)\exp(jkz) \quad (29)$$

where $r$ is the reflection coefficient. Hence, the light in the cavity can be written as,

$$\begin{aligned}E_g(x,y,z) &= E(x,y,z) + E_r(x,y,z) \\ &= E_0 \exp\left(-\dfrac{x^2+y^2}{w^2}\right)(1+r)\exp(jkz) + j2r\sin(kz)E_0\exp\left(-\dfrac{x^2+y^2}{w^2}\right)\end{aligned} \quad (30)$$

The first term is the travelling wave and the second term is a standing wave.

At the resonant condition of the Fabry-Perot cavity, the relation of the light field in the cavity and the transmitted light field from the cavity can be written as (13),

$$\frac{|E_g|^2}{|E_t|^2} = \frac{4n_0}{nT}\sin^2\left(\frac{\pi}{l}z\right) \tag{31}$$

where $E_t$ is the transmission field, $l$ is the length of the cavity, $n_0$ and $n$ are the refractive-index of the entrance and exit substrates respectively.

**(a) Planar graphene covered on the back mirror**

In case of planar graphene covered on the back mirror of the cavity, it is assumed that the light field in the cavity is a plane wave with Gaussian distribution. Thus we can obtain the nonlinear phase shifts induced by the graphene,

$$\Phi_{NL}(z,I) = \int k_0 \Delta n(I,z)dz = \frac{2\pi}{\lambda}\int n_2 I(x,y,z)dz \tag{32}$$

where $I(x,y,z)$ is the light intensity in the nonlinear graphene,

$$I(x,y,z) = \frac{n}{2z_0}|E_g|^2 \tag{33}$$

The transmission light is

$$I_t(x,y,z) = \frac{n}{2z_0}|E_t|^2 \tag{34}$$

Hence,

$$\frac{I(x,y,z)}{I_t(x,y,z)} = \frac{|E_g|^2}{|E_t|^2} = \frac{4n_0}{nT}\sin^2\left(\frac{\pi}{l}z\right) \tag{35}$$

The nonlinear phase shift is

$$\Phi_{NL}(z,I) = \frac{2\pi}{\lambda}\int n_2 I(x,y,z)dz = \frac{2\pi}{\lambda}\frac{4n_0}{nT}n_2 I_t \frac{l}{\pi}\int_{l-d}^{l}\sin^2\left(\frac{\pi}{l}z\right)d\left(\frac{\pi}{l}z\right)$$

$$= \frac{2\pi}{\lambda}\frac{4n_0}{nT}n_2 I_t \frac{l}{\pi}\left(\frac{\pi d}{2l} - \frac{1}{4}\sin\left(\frac{\pi}{l}2l\right) + \frac{1}{4}\sin\left[\frac{\pi}{l}2(l-d)\right]\right) \tag{36}$$

As the thickness of the graphene is very small, d<<l ,

$$\Phi_{NL}(z,I) \approx \frac{2\pi}{\lambda}\frac{4n_0}{n}\frac{d}{2}n_2\frac{I_t}{T} \tag{37}$$

Here, $T$ is the transmission of the back mirror. If we set $n_0 = n = 1$, hence

$$\Phi_{NL}(z,I) \approx \frac{2\pi}{\lambda} \frac{2I_t}{T} n_2 d \tag{38}$$

It can be found that the nonlinear phase shifts are inversely proportional to the transmission of the back-cavity mirror.

**(b) Graphene bubble on the back-cavity mirror**

Under the irradiation of the high intense light, the graphene bubble will be formed as discussed in previous sections. In order to simplify the simulation, we assume that the graphene bubble is semi-sphere with radius $R$, and light field is a plane wave with Gaussian distribution, then the nonlinear phase shift is associated with the transverse spatial distribution of light field

$$\Phi_{NL}(z,I) = \int k_0 \Delta n(I,z) dz = \frac{2\pi}{\lambda} \int n_2 I_{eff}(z) dz \tag{39}$$

here,

$$I_{eff}(z) = \frac{n}{2z_0} \frac{\int_{x,y} |E_g(x,y,z)|^2 dxdy}{\int_{x,y} dxdy} \tag{40}$$

At the resonant condition of the cavity,

$$I_{eff}(z) = \frac{n}{2z_0 \pi R^2} \int_{x,y} \frac{4n_0}{nT} \sin^2\left(\frac{\pi}{l}z\right) |E_t(x,y)|^2 dxdy \tag{41}$$

$$|E_t(x,y)|^2 = T|E_0|^2 \exp\left(-\frac{2(x^2+y^2)}{w^2}\right) \tag{42}$$

Then

$$I_{eff}(z) = \frac{n}{2z_0 \pi R^2} \frac{4n_0}{n} \sin^2\left(\frac{\pi}{l}z\right) |E_0|^2 \int_{x,y} \exp\left(-\frac{2(x^2+y^2)}{w^2}\right) dxdy \tag{43}$$

By using spherical coordinates,

$$\int_{x,y} \exp\left(-\frac{2(x^2+y^2)}{w^2}\right)dxdy = 2\pi R^2 \int_0^{\pi/2} \exp\left(-\frac{2R^2\sin^2\theta}{w^2}\right)\sin\theta d\theta$$

$$= -2\pi R^2 \exp\left(-\frac{2R^2}{w^2}\right)\int_0^{\pi/2} \exp\left(\frac{2R^2\cos^2\theta}{w^2}\right)d\cos\theta \qquad (44)$$

$$= -2\pi R^2 \exp\left(-\frac{2R^2}{w^2}\right)\frac{w}{\sqrt{2}R}\int_0^{\pi/2} \exp\left(\frac{2R^2\cos^2\theta}{w^2}\right)d\frac{\sqrt{2}R}{w}\cos\theta$$

If we set $x = \frac{\sqrt{2}R}{w}\cos\theta$,

$$\int_{x,y} \exp\left(-\frac{2(x^2+y^2)}{w^2}\right)dxdy = -2\pi R^2 \exp\left(-\frac{2R^2}{w^2}\right)\frac{w}{\sqrt{2}R}\int_{\frac{\sqrt{2}R}{w}}^{0} \exp(x^2)dx \qquad (45)$$

If we set $R \approx w$

$$\int_0^{\frac{\sqrt{2}R\pi}{2w}} \exp(x^2)dx \approx -3.344 \qquad (46)$$

Hence,

$$I_{eff}(z) = 6.688\frac{w}{\sqrt{2}R}\frac{n}{2z_0}\frac{4n_0}{n}|E_0|^2 \exp\left(-\frac{2R^2}{w^2}\right)\sin^2\left(\frac{\pi}{l}z\right) \qquad (47)$$

Finally, the nonlinear phase shift is

$$\Phi_{NL}(z,I) = \frac{2\pi}{\lambda}\int n_2 I_{eff}(z)dz =$$
$$\left(\frac{2\pi}{\lambda}n_2\right)6.688\frac{w}{\sqrt{2}R}\frac{n}{2z_0}\frac{4n_0}{n}|E_0|^2 \exp\left(-\frac{2R^2}{w^2}\right)\int_{l-d}^{l} \sin^2\left(\frac{\pi}{l}z\right)dz \qquad (48)$$

Here, $T$ is the transmission of the back mirror. If we set $n_0 = n = 1$, hence

$$\Phi_{NL}(z,I) \approx \left(\frac{2\pi}{\lambda}2n_2\right)2\frac{6.688w}{\sqrt{2}R}\frac{n}{2z_0}|E_0|^2 \exp\left(-\frac{2R^2}{w^2}\right)\int_{l-d}^{l} \sin^2\left(\frac{\pi}{l}z\right)dz \qquad (49)$$

Moreover, we suppose that the reflected light is a plane wave with Gaussian distribution, its average light intensity is

$$I_{t,eff} = \frac{\int_{x,y} I_t dxdy}{\int_{x,y} dxdy} = T\frac{n}{2z_0}|E_0|^2 \frac{w^2}{2R^2}\left(1-\exp\left(-\frac{2R^2}{w^2}\right)\right) \qquad (50)$$

Then we can written the relation between the nonlinear phase shifts and the transmitted light field

$$\Phi_{NL}(z,I) \approx \frac{\left(\frac{2\pi}{\lambda}\frac{2I_t}{T}n_2\right)2\frac{6.688}{\sqrt{2}}\exp\left(-\frac{2R^2}{w^2}\right)}{\frac{w}{2R}\left(1-\exp\left(-\frac{2R^2}{w^2}\right)\right)}\int_{l-d}^{l}\sin^2\left(\frac{\pi}{l}z\right)dz \qquad (51)$$

If we set $R \approx w$,

$$\Phi_{NL}(z,I) \approx 3\left(\frac{2\pi}{\lambda}\frac{2I_t}{T}n_2\right)\int_{l-d}^{l}\sin^2\left(\frac{\pi}{l}z\right)dz \qquad (52)$$

Here, the graphene forms a hemispherical surface, but there is still no change in the thickness, however, it is quite challenge to calculate $\int_{l-R}^{l-R+d}\sin^2\left(\frac{\pi}{l}z\right)dz$ in the propagation direction. Therefore, we adopt the approximate method and assume that the graphene is located between $z=l-R$ and $z=l-R+d$, then

$$\int_{l-R}^{l-R+d}\sin^2\left(\frac{\pi}{l}z\right)dz \approx \frac{l}{\pi}\left(\frac{\pi d}{2l} - \frac{1}{4}\sin\left(\frac{\pi}{l}2(l-R)\right) + \frac{1}{4}\sin\left[\frac{\pi}{l}2(l-R+d)\right]\right) \qquad (53)$$

If we set $R \approx l/4$, then

$$\int_{l-R}^{l-R+d}\sin^2\left(\frac{\pi}{l}z\right)dz \approx \frac{d}{2} + \frac{l}{\pi}\frac{1}{4}\left(\sin\frac{2\pi}{l}(l-R+d) - \sin\frac{2\pi}{l}(l-R)\right)$$

$$= \frac{d}{2} + \frac{l}{\pi}\frac{1}{4}\left(\sin\left(\frac{3}{2}\pi + \frac{2\pi}{l}d\right) - \sin\frac{3}{2}\pi\right) \qquad (54)$$

If we set $d = l/100$,

$$\int_{l-R}^{l-R+d}\sin^2\left(\frac{\pi}{l}z\right)dz \approx \frac{d}{2} + \frac{l}{\pi}\frac{1}{4}\left(\sin\left(\frac{3}{2}\pi + \frac{2\pi}{l}\frac{l}{100}\right) + 1\right)$$

$$\approx \frac{d}{2} + \frac{l}{\pi}\frac{1}{4}(-0.995 + 1) = \frac{d}{2} + 0.00004l \approx \frac{d}{2} \qquad (55)$$

As a result, the nonlinear phase shifts of the graphene bubble and planar graphene are $\Phi_{NL,bubble}(z,I) \approx \frac{2\pi}{\lambda}\frac{3I_t}{T}n_2 d$ and $\Phi_{NL,planar}(z,I) \approx \frac{2\pi}{\lambda}\frac{2I_t}{T}n_2 d$, respectively.

If we have follow parameters: T=0.01, $d = 1.0nm$, $I_t$=5×10$^{10}$ W/m$^2$, then we have $\Phi_{NL,bubble}(z,I) \approx 0.564\pi$ and $\Phi_{NL,planar}(z,I) \approx 0.376\pi$. The phase shift in graphene nanobubble is found to be nearly two times larger than that of planar graphene film.

**(6) The simulation of optical field of graphene bubble**

We use the finite-difference time-domain (FDTD) approach to simulate the focusing effect of the graphene bubble. The perfectly matched layer (PML) boundary condition have been applied to the X and Z direction, and the plane wave source with the wavelength of 532 nm is radiated from the top of the graphene bubble (along Z direction). The graphene bubble is considered as a sphere surface with a thickness of 1 nm and diameter of 750 nm. The size of the simulation region is 1800 nm × 1000 nm. The refractive-index of graphene is assumed to be $n_g = n_g' + \Delta n_g$, where $n_g' = 3$ is the linear refractive-index (*14*), $\Delta n_g = n_2 I_{eff}$ is the refractive index changes caused by the nonlinear effect, in which $I_{eff}$ is the effective light intensity in the cavity. In the simulation, the nonlinear refractive-index $n_2$ is obtained from previous literature reports (*15, 16*).

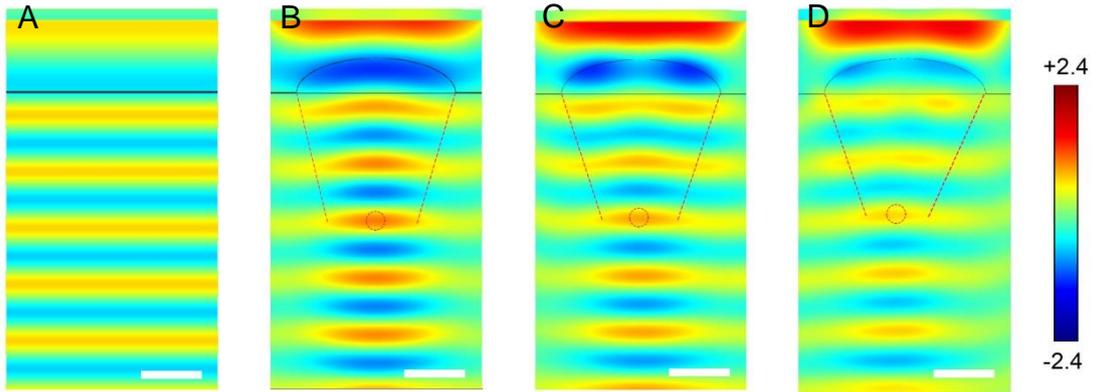

**Fig. S16. Simulation of optical field of flat graphene and graphene nanobubble.** **(A)** Simulated optical field of flat graphene on mirror substrate using FDTD (laser intensity of 1 ×10$^{10}$ W/m$^2$). **(B)** Simulated optical field of graphene bubble which has been just grown by a laser intensity of 1 ×10$^{10}$ W/m$^2$. **(C)** Simulated optical field on graphene bubble under a laser intensity of 5 ×10$^{11}$ W/m$^2$ (before the cavity is switched

on). **(D)** Simulated optical field on graphene bubble under a laser intensity of $1 \times 10^{12}$ W/m$^2$ (after the cavity is switched on). Scale bars in e-h: 300 nm. The black lines represent graphene film and the region below it refers to mirror substrate. The dashed lines in red indicate the focusing effect and the dashed circles in red show the center of focus points.

In this scenario, we consider the possibility of whether a curved graphene sheet can behave like an adaptive focus Kerr lens with the aid of finite-difference time-domain (FDTD) simulation (Fig. S16). In our experiments, water is possible to be sealed within graphene film and do not evaporate at elevated temperatures due to the impermeabilty of graphene *(17)*. The water layer thickness is more likely to be a factor of ten times smaller than the graphene nanobubble. For simplicity, this calculation assumes the graphene lens to be fully filled with water, but in reality such a scenario is unlikely.

We find that relatively large graphene bubbles can form under the focus laser beam with power density $\sim 1 \times 10^{10}$ W/m$^2$ (much higher than the saturation intensity of $1.3 \times 10^9$ W/m$^2$). In our simulations, a focused spot centered at 643 nm below the graphene bubble is already observable even though Kerr effect is not prominent at this power density, as illustrated in Fig. S16B. When the input power increases to the switch-on power ($2.7 \times 10^{11}$ W/m$^2$, corresponding to $I_{cavity}=5 \times 10^{11}$ W/m$^2$), we find that such graphene bubble can focus light beam into a spot centered at 615 nm below the mirror surface (Fig. S16C). Shortly after switching on, the power inside the cavity reaches $1 \times 10^{12}$ W/m$^2$. The light beam is strongly focused with even shorter focal length due to strong Kerr effect in terms of self-focusing, i.e., the center of focus point is shifted to 600 nm below the mirror surface, as shown in Fig. S16D. The beam waist of the focus point is found to be reduced from 500 nm (Fig. S16B) to 400 nm (Fig. S16C) and then to 260 nm (Fig. S16D). As a result, intensity-dependent phase change in graphene bubble is enhanced.

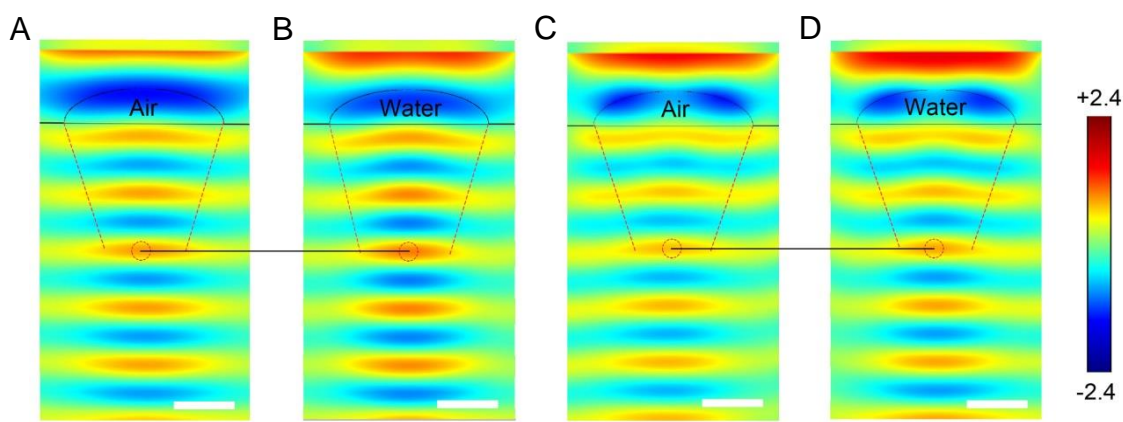

**Fig. S17. Simulation of optical field of graphene nanobubbles filled with air and water.** (**a**) Simulated optical field of graphene bubble filled with air under a laser intensity of $1 \times 10^{10}$ W/m$^2$. (**b**) Simulated optical field of graphene bubble filled with water under a laser intensity of $1 \times 10^{10}$ W/m$^2$. (**c**) Simulated optical field on graphene bubble filled with air under a laser intensity of $5 \times 10^{11}$ W/m$^2$). (**d**) Simulated optical field on graphene bubble filled with water under a laser intensity of $5 \times 10^{11}$ W/m$^2$. Scale bars in A-D: 300 nm. The black lines represent graphene film and the region below it refers to mirror substrate. The dashed lines in red indicate the focusing effect, the dashed circles in red show the center of focus points and the solid black lines suggest the same focal length.

In order to further verify whether the trapped material, i.e., water, will cause any nonlinear optical effect, we consider another optimal case in which graphene nanobubble is filled with dry air only. The simulation results are shown in Fig. S17. We can see that the optical field distribution caused by graphene bubble filled with air is almost the same as that by graphene bubble filled with water. Graphene bubble filled with air can also give similar self-focusing effect upon intense light illumination. The difference is that the optical field intensity at the focus point (Fig. S17, A and C) is slightly weaker than that of graphene bubble filled with water (Fig. S17, B and D), which is mainly due to the smaller refractive index of air in comparison with water (refractive index: 1.33). More importantly, it must be noted that the focal lengths of air-filled graphene bubble and water-filled graphene bubble are almost the same, which indicates the same nonlinear optical phase shift upon intense light illumination. This is explainable as water is normally not considered as an optical nonlinear matter, the incorporation of water will not contribute to the change in the nonlinear phase shifts which is expressed as $\Phi_{NL} = 4\pi n_2 I_{cavity} d / \lambda$ and correlated to $n_2$, instead it only slightly modifies the linear refractive index $n_o$ in $n(I_{cavity})=n_0 + n_2 I_{cavity}$. This result also agrees with previous experimental demonstration of optical bistability measurements on non-annealed and annealed graphene samples (in Section 8). Furthermore, one should consider the fact that graphene has very large third-order optical nonlinearity ($\chi^{(3)} \sim 10^{-7}$ esu) *(15,16)*, which is a few orders of magnitude higher than that of water ($\chi^{(3)} = 1.8 \times 10^{-14}$ esu). Therefore, we conclude that water locked in the graphene bubble will not cause observable nonlinear effect and the optical bistability is mainly due to graphene bubble.

**References**

1. A. C. Ferrari, D. M. Basko, *Nat. Nanotech.* **8**, 235 (2013).

2. A. Ferrari *et al.*, *Phys. Rev. Lett.* **97**, 187401 (2006).

3. J. Zabel *et al.*, *Nano Lett.* **12**, 617 (2012).

4. Z. H. Ni *et al.*, *Nano Lett.* **10**, 751 (2010).

5. X. Li *et al.*, *Science* **324**, 1312 (2009).

6. T. N. C. Venkatesan, City University of New York (1977).

7. H. Gibbs, *Optical bistability: Controlling light with light*.  (Academic Press, Inc., Orlando, Florida, 1985), vol. 1, pp. 481.

8. Q. Bao *et al.*, *Adv. Funct. Mater.* **19**, 3077 (2009).

9. M. Born, E. Wolf, *Cambridge, England*,  (1999).

10. D. Miller, *Quantum Electronics, IEEE Journal of* **17**, 306 (1981).

11. A. Szoke, V. Daneu, J. Goldhar, N. Kurnit, *Appl. Phys. Lett.* **15**, 376 (1969).

12. F. Wang *et al.*, *Science* **320**, 206 (2008).

13. T. Hattori, N. Tsurumachi, H. Nakatsuka, *JOSA B* **14**, 348 (1997).

14. C. Casiraghi *et al.*, *Nano Lett.* **7**, 2711 (2007).

15. E. Hendry, P. J. Hale, J. Moger, A. K. Savchenko, S. A. Mikhailov, *Phys. Rev. Lett.* **105**, 097401 (2010).

16. H. Zhang *et al.*, *Opt. Lett.* **37**, 1856 (2012).

17.  C. H. Y. X. Lim *et al.*, *Nat. Comm.* **4**, 1556 (2013).